\documentclass[prd,showpacs,preprint]{revtex4}
\usepackage{amssymb}
\usepackage{amsmath}
\usepackage{graphicx}
\usepackage{multirow}
\usepackage{subfigure}
\usepackage{float}

\begin{document}

\title{Next-to-leading order QCD corrections to $tZ$ associated  production via the flavor-changing
neutral-current couplings at hadron colliders}
\author{Bo Hua Li}
\author{Yue Zhang}
\author{Chong Sheng Li}
\email{csli@pku.edu.cn}
\author{Jun Gao}
\author{Hua Xing Zhu}
\affiliation{Department of Physics and State Key Laboratory of
Nuclear Physics and Technology, Peking University, Beijing 100871,
China}

\date{\today}


\begin{abstract}
We present the complete next-to-leading order (NLO) QCD  corrections to $tZ$
associated production induced by the model-independent $tqg$ and $tqZ$ flavor-changing
neutral-current couplings at hadron colliders, respectively.
Our results show that, for the $tuZ$ coupling the NLO QCD corrections can enhance the
total cross sections by about 60\% and 42\%, and for the $tcZ$
coupling by about 51\% and 43\% at the Tevatron and LHC,
respectively. The NLO corrections, for the $tug$ couplings,
can enhance the total cross sections by about 27\%, and by about 42\% for the $tcg$ coupling at the LHC.
 We also consider the mixing effects between the $tqg$ and $tqZ$
couplings for this process, which can either be large or small depending on the values of the
 anomalous couplings. Besides, the NLO corrections reduce the dependence
of the total cross sections on the renormalization or factorization
scale significantly, which lead to increased confidence on the
theoretical predictions. And we also evaluate the NLO corrections to several important kinematic distributions.

\end{abstract}

\maketitle

\section{Introduction}\label{s1}
The mass of the top quark is close to the electroweak(EW) symmetry
breaking scale, and thus its decay and production at colliders are
very important for the probe of the EW breaking mechanism and new
physics beyond the standard model (SM). Direct evidence of new
physics at TeV scale may be not easy to find, while indirect
evidence, such as modification of SM predictions originated from new
physics interaction, is important as well. A good consideration is
to investigate the single top quark production process via the
anomalous flavor-changing neutral-current (FCNC) couplings. Within
the SM, the FCNC couplings can only occur at the loop level, which
are further suppressed by the GIM mechanism~\cite{Glashow:1970gm}.
On the other hand, some new physics
models~\cite{AguilarSaavedra:2004wm,Larios:2006pb} such as the two Higgs doublet
models~\cite{Cheng:1987rs}, supersymmetric models~\cite{Li:1993mg},
extra dimensions models~\cite{Davoudiasl:2001uj}, may enhance these
FCNC couplings to observable level.
 As the coming Large Hadron Collider (LHC)
will produce abundant top quark events (about $10^8$ per year), even
in the initial low luminosity run ($\sim 10{\rm \ fb}^{-1}$/year)
$8\times10^6$ top quark pairs and $3\times10^6$ single top quarks
will be produced yearly,
one may anticipate the discovery of the first
hint of new physics by observing the FCNC couplings in the top quark
sector.

In general, any new physics at a high energy scale $\Lambda$ can be
described by an effective Lagrangian containing higher dimensional
SM gauge invariant operators~\cite{Buchmuller:1985jz}. For the new
physics induced top-quark FCNC couplings related to the gluon and $Z$
boson, respectively, they can be incorporated into the dimension
five effective operators as listed below~\cite{Beneke:2000hk},
\begin{equation}\label{lag}
-g_s\sum_{q=u,c}\frac{\kappa^g_{tq}}{\Lambda}\bar{q}\sigma^{\mu\nu} T^a (f^g_{tq}+ih^g_{tq}\gamma_5)
t G^a_{\mu\nu}-\frac{e}{\text{sin2$\theta $}_W}\sum_{q=u,c}\frac{\kappa^z_{tq}}{\Lambda}\bar{q}\sigma^{\mu\nu}
(f^Z_{tq}+ih^Z_{tq}\gamma_5) t Z_{\mu\nu}+H.c.,
\end{equation}
where $\Lambda$ is the new physics scale, $T^a$ are the Gell-Mann
matrices, $G^a_{\mu\nu}$ and $Z_{\mu\nu}$ are the field strength
tensors of the gluon and $Z$ boson, respectively, $\kappa^z_{tq}\
(q=u,c)$ and $\kappa^g_{tq}\ (q=u,c)$ are real coefficients that
define the strength of the couplings. And $\text{$\theta $}_W$ is
the weak-mixing angle, while $f^V_{\mathrm{tq}}$ and
$h^V_{\mathrm{tq}}$ are complex numbers satisfying
$|f^V_{\mathrm{tq}}|^2+|h^V_{\mathrm{tq}}|^2=1$ with $V=Z,g$ and $q
= u, c$.

The CDF collaboration has set 95\% confidence level (CL) limits on the
branching ratio $BR(t \to qZ) < 0.037$~\cite{:2008aaa}, which corresponds to $\kappa^Z_{tq}/\Lambda <0.908\rm{TeV}^{-1}$
based on the theoretical predictions of $t \to qZ$ at the NLO level in QCD~\cite{Zhang:2008yn,Zhang:2010bm,Drobnak:2010wh,Fox:2007in}.
The D0 collaboration also provides a more stringent constrains, $BR(t \to qZ) < 0.032$ at a
 95\% confidence level~\cite{Abazov:2011qf}, which corresponds to $\kappa^Z_{tq}/\Lambda <0.845\rm{TeV}^{-1}$.
Currently the most stringent experimental constraints for the $tqg$
anomalous couplings are $\kappa^g_{tu}/\Lambda\leq 0.013\ {\rm
TeV^{-1}}$ and $\kappa^g_{tc}/\Lambda\leq 0.057\ {\rm TeV^{-1}}$,
given by the D0 collaboration~\cite{Abazov:2010qk}, and
$\kappa^g_{tu}/\Lambda\leq 0.018\ {\rm
TeV^{-1}}$ and $\kappa^g_{tc}/\Lambda\leq 0.069\ {\rm TeV^{-1}}$ given by the CDF
collaboration~\cite{Aaltonen:2008qr}, both based on the measurements of the
FCNC single top production using the theoretical predictions, including the
NLO QCD corrections~\cite{Liu:2005dp,Gao:2009rf} and resummation effects~\cite{Yang:2006gs}, respectively.

Since the observation of $pp\rightarrow tZ$ is a
clear signal of top flavor violation, and we do not know  which
type of new physics will be responsible
for a future deviation from the SM predictions, it is necessary to
study this process in a model-independent way.
There are already several literatures~\cite{AguilarSaavedra:2004wmetc,Kidonakis:2003} discussing this process
using the effective Lagrangian in Eq.~(\ref{lag}). However, they were either
based on the LO calculations~\cite{AguilarSaavedra:2004wmetc}, or the NLO QCD effects are not
completely calculated~\cite{Kidonakis:2003}. So it is necessary to present a complete NLO QCD
corrections to the above process, which is not only mandatory for
matching the expected experimental accuracy at hadron colliders, but
is also important for a consistent treatment of both the top quark
production and decay via the FCNC couplings by experiments.
In this paper, we present the complete NLO QCD corrections to
$tZ$ associated production
via $tqZ$ and $tqg$ FCNC couplings with their mixing effects at hadron colliders.

The arrangement of this paper is as follows. In Sec.~\ref{s2} we
show the LO results for the process induced by $tqZ$ FCNC couplings.
In Sec.~\ref{s3}, we present the details of the
NLO calculations, including the virtual and real corrections. We
discuss the process induced by $tqg$ FCNC couplings and the mixing effects in Sec.~\ref{s4},
 Sec.~\ref{s5} contains the numerical results,
and Section~\ref{s6} is a brief summary.

\section{Leading order results}\label{s2}
\begin{figure}[h]
\begin{center}
\scalebox{0.7}{\includegraphics*{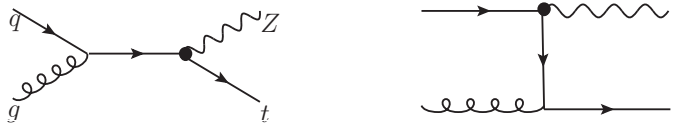}}
\caption[]{\label{treekappaZ}The LO Feynman diagrams for the single top
quark production via the FCNC couplings without operator mixing.}
\end{center}
\end{figure}

At hardron colliders, there is only one subprocess
that contributes to the $tZ$ associated
production at the LO via the electroweak FCNC couplings,
$\kappa^{Z}_{tq}$:
\begin{equation}
g \ q \longrightarrow t \ Z,
\end{equation}
where $q$ is either $u$ quark or $c$ quark. The corresponding
Feynman diagrams are shown in Fig.~\ref{treekappaZ}.

After sum over the spins and colors of the outgoing particles and
average over the spins and colors of the incoming particles,
 the LO squared amplitude is
\begin{eqnarray}\label{eq1}
\overline{|M^B|^2}_{gq}(s,t)&=& \frac{32 \pi ^2
 \alpha  \alpha _s \kappa _z^2}{
3sin(2\theta_W)^2 \Lambda ^2 s \left(t-m_t^2\right)^2}
(2 m_t^8-(3 m_z^2+4 s+2 t) m_t^6+ \nonumber \\ &&
(2 m_z^4-(2 s+t) m_z^2+2(s^2+4 t s+t^2)) m_t^4+(2 m_z^6-4 t m_z^4 \nonumber \\ &&
+(s^2+6 t s+5 t^2)m_z^2-2 t(3 s^2+6 t s+t^2)) m_t^2+t (-2 m_z^6+2 \nonumber \\ &&
(s+t) m_z^4 -(s+t)^2 m_z^2+4 s t (s+t))),
\end{eqnarray}
where $m_t$ is the top quark mass, and
$m_z$ is the $Z$ boson mass, $s$, $t$, and
$u$ are the Mandelstam variables, which are defined as
\begin{equation}
s=(p_1+p_2)^2,\ \ t=(p_1-p_3)^2,\ \ u=(p_1-p_4)^2.
\end{equation}
After the phase space integration, the LO partonic cross sections
are given by
\begin{equation}
\hat \sigma^B_{ab}=\frac{1}{2\hat s}\int d\Gamma
\overline{|M^B|^2}_{ab}.
\end{equation}
The LO total cross section at hadron colliders is obtained by
convoluting the partonic cross section with the parton distribution
functions (PDFs) $G_{i/P}$ for the proton (antiproton)
\begin{equation}
\sigma^B=\sum_{ab}\int dx_1
dx_2\left[G_{a/P_1}(x_1,\mu_f)G_{b/P_2}(x_2,\mu_f)\hat
\sigma^B_{ab}\right],
\end{equation}
where $\mu_f$ is the factorization scale.

\section{Next-to-leading order QCD corrections}\label{s3}

\begin{figure}[h]
\begin{center}
\scalebox{0.5}{\includegraphics*[60,324][1510,794]{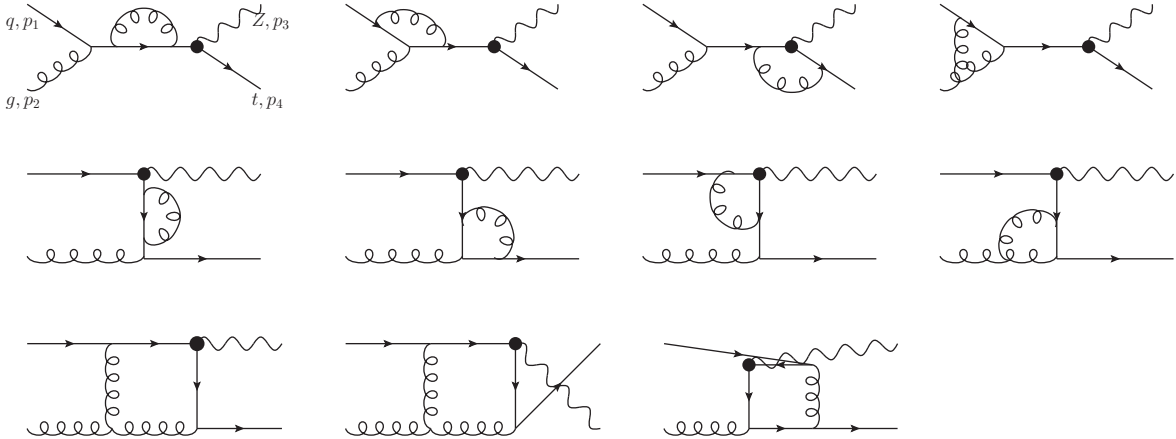}}
\caption[]{\label{f2}1-loop Feynman diagrams for the single top
quark production via the FCNC couplings without operator mixing.}
\end{center}
\end{figure}

In this section, we present our calculations for the NLO QCD
correcions to the $tZ$ associated production via
the electroweak FCNC couplings.
The NLO corrections include both the virtual and the real corrections
with the Feynman diagrams shown in Figs.~\ref{f2}-\ref{f3}, which are generated
with FeynArts~\cite{Hahn:2000kx}, and calculated with FormCalc~\cite{Hahn:2010zi}.
We use the dimensional regularization (DREG) scheme~\cite{'tHooft:1972fi} with naive $\gamma_5$ prescription in
$n=4-2\epsilon$ dimensions to regularize all the divergences.
Moreover, for the real corrections, we use the two-cutoff phase space slicing
method~\cite{Harris:2001sx} to separate the infrared(IR) divergences.
\subsection{Virtual corrections}

The virtual corrections contains both UV and IR divergences, with the UV
divergences renormalized by introducing counterterms.
For the external fields, we define all the renormalization constants using
the on-shell subtraction scheme
\begin{eqnarray}\label{deltaZ}
\delta Z_2^{(g)}&=&-\frac{\alpha_s}{2\pi}C_{\epsilon}\left(
\frac{n_f}{3}-\frac{5}{2}\right)\left(\frac{1}{\epsilon_{UV}}-
\frac{1}{\epsilon_{IR}}\right)-\frac{\alpha_s}{6\pi}C_{\epsilon}
\frac{1}{\epsilon_{UV}}, \nonumber \\
\delta Z_2^{(q)}&=&-\frac{\alpha_s}{3\pi}C_{\epsilon}\left(\frac{1}
{\epsilon_{UV}}-\frac{1}{\epsilon_{IR}}\right), \nonumber \\
\delta Z_2^{(t)}&=&-\frac{\alpha_s}{3\pi}C_{\epsilon}\left(\frac{1}
{\epsilon_{UV}}+\frac{2}{\epsilon_{IR}}+4\right), \nonumber \\
\frac{\delta m_t}{m_t}&=&-\frac{\alpha_s}{3\pi}C_{\epsilon}\left(\frac
{3}{\epsilon_{UV}}+4\right),
\end{eqnarray}
where $C_{\epsilon}=\Gamma(1+\epsilon)[(4\pi\mu_r^2)/m_t^2]
^{\epsilon}$ and $n_f=5$. For the renormalization of the strong coupling constant $g_s$,
and the FCNC couplings $\delta
Z_{\kappa^g_{tq}/\Lambda}$, we use
the $\rm \overline{MS}$ scheme~\cite{Zhang:2008yn}:
\begin{eqnarray}\label{deltaKz}
\delta Z_{g_s}&=&\frac{\alpha_s}{4\pi}\Gamma(1+\epsilon)
(4\pi)^{\epsilon}\left({n_f\over 3}-{11\over 2}\right){1\over
\epsilon_{UV}}+{\alpha_s\over {12\pi}}C_{\epsilon}{1\over
\epsilon_{UV}}, \nonumber \\
\delta
Z_{\kappa^Z_{tq}/\Lambda}&=&\frac{\alpha_s}{3\pi}\Gamma(1+\epsilon)(4\pi)
^{\epsilon}{1\over \epsilon_{UV}},
\end{eqnarray}
and the running of the FCNC couplings are given
by~\cite{Zhang:2008yn}
\begin{equation}
\frac{\kappa^Z_{tq}(\mu)}{\Lambda}=\frac{\kappa^Z_{tq}(\mu')}{\Lambda}
\left(\frac{\alpha_s(\mu')}{\alpha_s(\mu)}\right)^{4/ (3\beta_0)},
\end{equation}
with $\beta_0=11-2n_f/3$.

All the UV divergences cancel each other,
leaving the remaining IR divergences and the finite
terms. Because of the limited space, we do not shown the lengthy explicit
expressions of the virtual corrections here. The IR divergence of the virtual
corrections to the partonic total cross section can be factorized as~\cite{Denner:1991kt,Ellis:2007qk}
\begin{eqnarray}\label{eq2}
 \hat\sigma^{Loop}_{IR}&=&-\frac{\alpha _s}{6 \pi }D_{\epsilon}
\Big\{\frac{13}{\epsilon _{\text{IR}}^2}+\left(4 \text{ln}(\frac{s}{m_t^2})
+\text{ln}(\frac{m_t^2-u}{m_t^2})-9 \text{ln}
(\frac{m_t^2-t}{m_t^2})+\frac{43}{2}\right)\frac{1}{\epsilon
_{\text{IR}}}\Big\}\hat\sigma^B,
\end{eqnarray}
where $D_{\epsilon}=[(4\pi\mu_r^2)/s]^
{\epsilon}/\Gamma(1-\epsilon)$.
In order to cancel these divergences, we need to extract the
IR divergences in the real corrections, which will be shown in the following subsection.

\subsection{Real corrections}\label{ss1}

\begin{figure}[h]
      \begin{center}
     \scalebox{0.5}{\includegraphics*[25,150][1980,750]{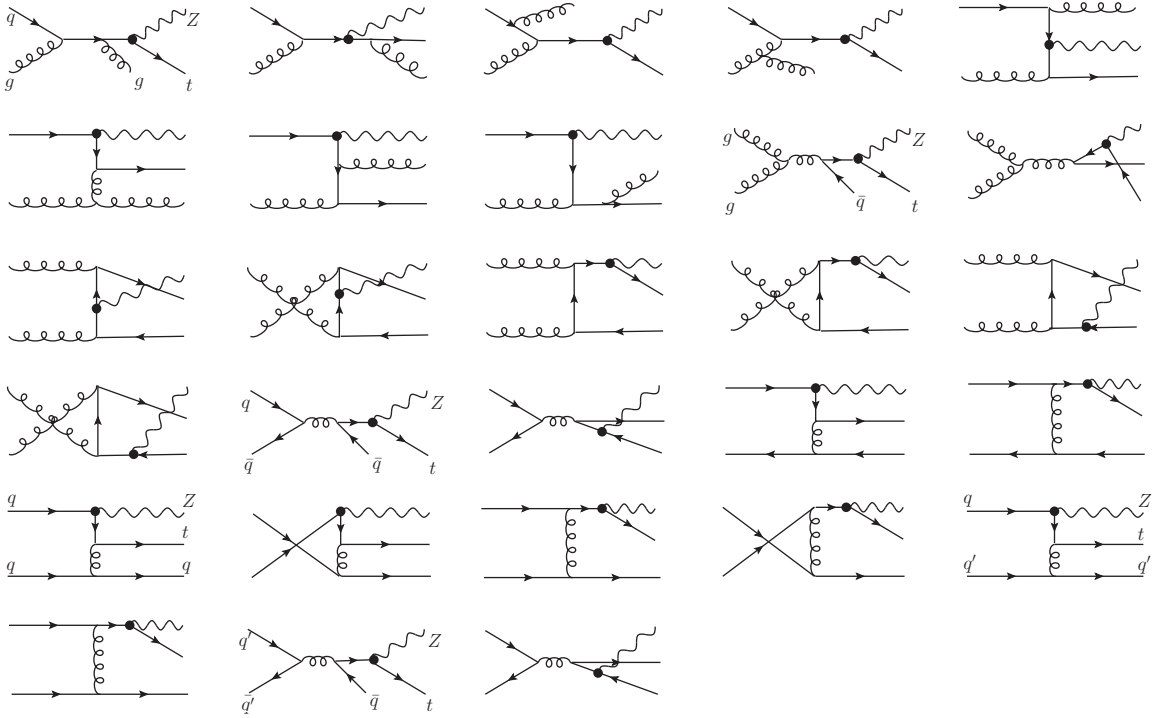}}
      \end{center}
    \caption[]{\label{f3}Feynman diagrams of the real corrections for the single top
quark production via the FCNC couplings without operator mixing.}
\end{figure}

The real corrections consist of the radiations of an
additional gluon $g\ q\longrightarrow t\ Z\ g$, or massless quark(anti) in the final states,
$\ g\ g\longrightarrow t\ \bar q\ Z,\ q \ q(\bar{q},q') \longrightarrow t \
q(\bar{q},q')\ Z,\ q'\bar{q'} \longrightarrow t \
\bar{q}\ Z  $ as shown in Fig.~\ref{f3}. It should be noted that in our NLO
calculations of the process induced by $tqZ$ couplings, we did not include the contributions from the SM
on-shell production of the top quark pair with subsequent rare decay of
one top quark, $pp(\bar p) \to t\bar t\to t +\bar q+ Z$, and also the corresponding interference
terms, following the diagram removal scheme proposed in reference~\cite{Frixione:2008yi}.
This procedure does violate the gauge invariance because certain diagrams
are removed, but the influence to the numerical results is small as show in~\cite{Frixione:2008yi}.
We have also crosschecked the numerical results by using another method preserving gauge invariance~\cite{Tait:1999cf},
where an invariant mass cut of the Z boson and light quark is adopted. And the results of the invariant mass cut method agree well with
the ones of the diagram removal method if
we require the invariant mass to be out of the range of $\pm17\Gamma_t$ of the top quark mass, where $\Gamma_t$ is the width of top quark.

\subsubsection{Real gluon emission}

For real gluon emission, the phase space integration contains both soft
and collinear singularities. We adopt the two-cutoff phase space slicing method to
isolate all the IR singularities~\cite{Harris:2001sx}, which introduces two small parameters $\delta_s$ and
$\delta_c$ to divide the phase space into three parts. The soft
cutoff $\delta_s$ separates the phase space into the soft region
 $E_5\leq\delta_s\sqrt{s}/2$ and the hard region,
\begin{equation}
\hat{\sigma }^R = \hat{\sigma }^{\text{H}}+\hat{\sigma }^S,
\end{equation}
furthermore, the hard piece can be divided into two sub-regions by $\delta_c$,
 \begin{equation}
\hat{\sigma }^{\text{H}} = \hat{\sigma }^{\overline{\text{HC}}}+\hat{\sigma }^{\text{HC}}.
\end{equation}

The hard noncollinear part $\hat{\sigma }^{\overline{\text{HC}}}$ is finite
and the phase space integration can be calculated numerically.
For the soft region, in the limit that the energy of the emitted gluon becomes small,
i.e. $E_5\leq \delta_s\sqrt{s}/2$, the amplitude squared
$\overline{\sum}|M(qg) \to t Z +g)|^2$ can be
factorized into the Born amplitude squared times an eikonal factor
$\Phi_{\text{eik}}$

\begin{equation}
\overline{\sum}|M(qg) \to t Z +g)|^2
\stackrel{\text{soft}}{\longrightarrow}
(4\pi\alpha_s\mu_r^{2\epsilon}) \overline{\sum}|M^B|^2
\Phi_{\text{eik}},
\end{equation}
where the eikonal factor is given by
\begin{eqnarray}
\Phi_{eik}&=&\frac{C_A}{2}\frac{s}{(p_{1}\cdot p_{5})(p_{2}\cdot p_{5})}-\frac{1}{2C_A}\frac{m_t^2-u}{(p_{1}\cdot p_{5})(p_{4}\cdot p_{5})} \nonumber\\
&&+\frac{C_A}{2}\frac{m_t^2-t}{(p_{2}\cdot p_{5})(p_{4}\cdot
p_{5})}-C_F\frac{m_t^2}{(p_{4}\cdot p_{5})^2},
\end{eqnarray}
where $C_A=3, C_F=\frac{4}{3}$. Moreover, the three-body phase space
in the soft limit can also be factorized
\begin{equation}
d\Gamma_3(qg \to tZ
+g)\stackrel{\text{soft}}{\longrightarrow}d\Gamma_2(qg \to
tZ)dS.
\end{equation}
Here $dS$ is the integration over the phase space of the soft gluon
which is given by\cite{Harris:2001sx}
\begin{equation}
dS = \frac{1}{2(2\pi)^{3-2\epsilon}} \int^{\delta_s \sqrt{s}/2}_0
dE_5E_5^{1-2\epsilon}d\Omega_{2-2\epsilon}.
\end{equation}
The parton level cross section in the soft region can be
expressed as
\begin{equation}
\label{sigma:soft1} \hat{\sigma}^S =
(4\pi\alpha_s\mu^{2\epsilon}_r)\int
d\Gamma_2\overline{\sum}|M^B|^2\int dS \Phi_{\text{eik}}.
\end{equation}
After the integration over the soft gluon phase space\cite{Harris:2001sx},
 Eq.(\ref{sigma:soft1}) becomes
\begin{equation}
\hat{\sigma}^S = \hat{\sigma}^B
\left[\frac{\alpha_s}{2\pi}\frac{\Gamma(1-\epsilon)}{\Gamma(1-2\epsilon)
}\left(\frac{4\pi\mu^2_r}{s}\right)^{\epsilon}\right]\left(\frac{A^s_2}
{\epsilon^2}+\frac{A^s_1}{\epsilon}+A^s_0\right),
\end{equation}
with
\begin{eqnarray}
A^s_2&=&\frac{13}{6 \pi}, \nonumber \\
A^s_1&=&\frac{1}{6 \pi} \{-26 \text{ln} (\text{$\delta_s$})+4
\text{ln}
   (\frac{s}{m_t^2})-\text{ln}
   (\frac{m_t^2}{m_t^2-u})+9 \text{ln}(
   \frac{m_t^2}{m_t^2-t})+4\}
, \nonumber \\
A^s_0&=&\frac{1}{12 \pi }\big\{52 \text{ln} ^2(\text{$\delta_s$})-2
\Big[\text{ln} (\frac{\left(m_t^2-t\right)^2}{s m_t^2})-9 \text{ln}
(\frac{\left(-m_z^2+s+t\right)^2}{s m_t^2})+8\Big] \text{ln}
   (\text{$\delta_s$})  \nonumber \\ &&
   +A+9 B_--B_+\Big\},
\end{eqnarray}
where A and $B_\pm$, are given in Appendix.
\\

In the hard collinear region, $E_5> \delta_s\sqrt{s}/2$ and
$-\delta_c s< t_{i5} < 0$, the emitted hard gluon is collinear to
one of the incoming partons. As a consequence of the factorization
theorem\cite{Collins:1985ue, Bodwin:1984hc} the matrix element
squared for $qg \rightarrow tZ +g$ can be factorized into the
product of the Born amplitude squared and the Altarelli-Parisi
splitting function
\begin{equation}
\overline{\sum}|M(qg \rightarrow tZ +
g)|^2\stackrel{\text{collinear}}
{\longrightarrow}(4\pi\alpha_s\mu^{2\epsilon}_r)\overline{\sum}|M^B|^2\left(\frac{-2P_{qq}(z,\epsilon)}{zt_{15}}
+\frac{-2P_{gg}(z,\epsilon)}{zt_{25}}\right),
\end{equation}
where $z$ denotes the fraction of the momentum of the incoming
parton carried by $q(g)$ with the emitted gluon taking a fraction
$(1-z)$, and the unregulated Altarelli-Parisi splitting functions
are written explicitly as~\cite{Harris:2001sx}
\begin{eqnarray}
P_{qq}(z,\epsilon) &=& C_{F}\Big(\frac{1+z^{2}}{1-z}-\epsilon(1-z)\Big), \nonumber \\
P_{gg}(z,\epsilon) &=& 2N\Big(\frac{z}{1-z}+\frac{1-z}{z}+z(1-z)\Big).
\end{eqnarray}

Moreover, the three-body phase space can also be factorized in the collinear limit,
for example, in the limit $-\delta_c s < t_{15} < 0$ it has the following form\cite{Harris:2001sx}
\begin{equation}
d\Gamma_3(qg \rightarrow tZ +
g)\stackrel{\text{collinear}}{\longrightarrow}d\Gamma_2(qg
\rightarrow tZ; s^{\prime} = zs)
\frac{(4\pi)^{\epsilon}}{16\pi^2\Gamma(1-\epsilon)}dzdt_{15}[-(1-z)t_{15}]^{-\epsilon}.
\end{equation}
Thus, after convoluting with the PDFs, the three-body cross section
in the hard collinear region is given by\cite{Harris:2001sx}
\begin{eqnarray}
d\sigma^{HC} & = & d\hat{\sigma}^B
\left[\frac{\alpha_s}{2\pi}\frac{\Gamma(1-\epsilon)}{\Gamma(1-2\epsilon)}
\left(\frac{4\pi\mu^2_r}{s}\right)^{\epsilon}\right](-\frac{1}{\epsilon})
\delta_c^{-\epsilon}\left[P_{qq}(z,\epsilon)G_{q/p}(x_1/z)G_{g/p}(x_2)
\right.\nonumber\\&& \left.+
P_{gg}(z,\epsilon)G_{g/p}(x_1/z)G_{q/p}(x_2) +
(x_1 \leftrightarrow x_2)\right]
\frac{dz}{z}\left(\frac{1-z}{z}\right)^{-\epsilon}dx_1dx_2.
\end{eqnarray}
where $G_{q(g)/p}(x)$ is the bare PDF.

\subsubsection{Massless (anti)quark emission}

In addition to the real gluon emission, a second set of real emission
corrections to the inclusive cross section for $pp\rightarrow tZ$
 at NLO involves the processes with an additional massless
$q(\bar q)$ in the final state. Since the contributions from real massless $q(\bar q)$ emission
contain initial state collinear singularities we need to use
the two cutoff phase space slicing method \cite{Harris:2001sx} to
isolate these collinear divergences. The
cross sections for the processes with an additional massless
$q(\bar{q})$ in the final state can be expressed as
\begin{eqnarray}
\label{sigma:nc} d\sigma^{add} & = &
\sum_{(\alpha=q,\bar{q},q')} \Big\{\hat{\sigma}^{\overline{C}}
(q \alpha \rightarrow tZ +
q(\bar{q}))G_{q/p}(x_1)G_{\alpha/p}(x_2)+ \nonumber\\ && d\hat{\sigma}^B
\left[\frac{\alpha_s}{2\pi}\frac{\Gamma(1-\epsilon)}{\Gamma(1-2\epsilon)}
\left(\frac{4\pi\mu^2_r}{s}\right)^{\epsilon}\right]
(-\frac{1}{\epsilon}) \delta_c^{-\epsilon}
P_{g\alpha}(z,\epsilon)G_{q/p}(x_1/z)G_{\alpha/p}(x_2) \nonumber\\
&& \frac{dz}{z}\left(\frac{1-z}{z}\right)^{-\epsilon}+ (x_1
\leftrightarrow x_2)\Big\}dx_1dx_2+ \Big\{\hat{\sigma}^{\overline{C}}
(gg \rightarrow tZ +\bar{q})G_{g/p}(x_1)G_{g/p}(x_2)+ \nonumber\\ && d\hat{\sigma}^B
\left[\frac{\alpha_s}{2\pi}\frac{\Gamma(1-\epsilon)}{\Gamma(1-2\epsilon)}
\left(\frac{4\pi\mu^2_r}{s}\right)^{\epsilon}\right]
(-\frac{1}{\epsilon}) \delta_c^{-\epsilon}
P_{qg}(z,\epsilon)G_{g/p}(x_1/z)G_{g/p}(x_2) \nonumber\\
&& \frac{dz}{z}\left(\frac{1-z}{z}\right)^{-\epsilon}+ (x_1
\leftrightarrow x_2)\Big\}dx_1dx_2,
\end{eqnarray}
where
\begin{eqnarray}
P_{qg}(z,\epsilon) & = & P_{\bar{q}g}(z) =
\frac{1}{2}[z^2+(1-z)^2]-z(1-z)\epsilon,\nonumber\\
P_{gq}(z,\epsilon) & = & P_{g\bar{q}}(z) =
C_F[\frac{z}{1+(1-z)^2}-z\epsilon].
\end{eqnarray}
The $\hat{\sigma}^{\overline{C}}$ terms in Eq. (\ref{sigma:nc})
represents the noncollinear cross sections for the $q(\bar{q},q,q')$ and $gg$
initiated processes which can be written in the form

\begin{eqnarray}
\hat{\sigma}^C=\frac{1}{2s}\Big\{\sum_{(\alpha=q,\bar{q},q')}
|M(q(\bar{q},q,q'))\stackrel{\text{collinear}}\longrightarrow tZ+(\bar{q},q,q')|^2+ |M(gg)
\stackrel{\text{collinear}}\longrightarrow tZ+\bar{q})|^2\Big\}d\bar{\Gamma_3},
\end{eqnarray}
where $d\bar{\Gamma}_3$ is the three-body phase space in the
noncollinear region. The other terms in Eq. (\ref{sigma:nc}) are the
collinear singular cross sections.

\subsubsection{Mass factorization}
After adding the renormalized virtual corrections and the real
corrections, the parton level  cross sections still contain collinear
divergences which can be absorbed into a redefinition of the PDFs at the
NLO, namely through mass factorization\cite{Altarelli:1979ub,
Collins:1989gx}. This procedure, in practice, means that first we
convolute the partonic cross section with the bare PDF
$G_{\alpha/p}(x)$ and then use the renormalized PDF
$G_{\alpha/p}(x,\mu_f)$ to replace $G_{\alpha/p}(x)$. In the
$\overline{\text{MS}}$ convention the scale-dependent PDF
$G_{\alpha/p}(x,\mu_f)$ is given by \cite{Harris:2001sx}
\begin{eqnarray}
\label{modifiedPDF} G_{\alpha/p}(x,\mu_f) & = & G_{\alpha/p}(x) +
\sum_{\beta}\left(-\frac{1}{\epsilon}\right)\left[
\frac{\alpha_s}{2\pi}\frac{\Gamma(1-\epsilon)}{\Gamma(1-2\epsilon)}
\times \left(\frac{4\pi\mu^2_r}{\mu_f^2}\right)^{\epsilon}\right]\nonumber\\
&& \times \int_x^1 \frac{dz}{z} P_{\alpha\beta}(z) G_{\beta/p}(x/z).
\end{eqnarray}
Then the $\mathcal{O}
(\alpha_s)$ expression for the remaining collinear contribution
can be written in the following form:
\begin{eqnarray}
&& d\sigma^{coll}=  d\hat{\sigma}^B\bigg[\frac{\alpha_s}{2\pi}
\frac{\Gamma(1-\epsilon)} {\Gamma(1-2\epsilon)}
\bigg(\frac{4\pi\mu^2_r}{s}\bigg)^\epsilon \bigg]
\{\tilde{G}_{q/p}(x_1,\mu_f) G_{g/p}(x_2,\mu_f) +
G_{q/p}(x_1,\mu_f) \tilde{G}_{g/p}(x_2,\mu_f) \nonumber
\\ && \hspace{1.2cm}
+\sum_{\alpha=q,g}\bigg[\frac{A_1^{sc}(\alpha\rightarrow
\alpha g)}{\epsilon} +A_0^{sc}(\alpha\rightarrow \alpha
g)\bigg]G_{q/p}(x_1,\mu_f) G_{\bar{q}/p}(x_2,\mu_f) \nonumber
\\ && \hspace{1.2cm}
+(x_1\leftrightarrow x_2)\} dx_1dx_2,\label{11}
\end{eqnarray}
where
\begin{eqnarray}
A_0^{sc}&=&A_1^{sc}\ln(\frac{s}{\mu_f^2}), \\
A_1^{sc}(q\rightarrow qg)&=&C_F(2\ln\delta_s +3/2), \\
A_1^{sc}(g\rightarrow gg)&=&2N\ln\delta_s + (11N-2n_f)/6, \\
\tilde{G}_{\alpha/p}(x,\mu_f)&=&\sum_{\beta,\alpha}\int_x^{1-
\delta_s\delta_{\alpha\beta}} \frac{dy}{y}
G_{\beta/p}(x/y,\mu_f)\tilde{P}_{\alpha\beta}(y),
\end{eqnarray}
with
\begin{eqnarray}
\tilde{P}_{\alpha\beta}(y)=P_{\alpha\beta}(y) \ln(\delta_c
\frac{1-y}{y} \frac{s}{\mu_f^2}) -P_{\alpha\beta}'(y),
\end{eqnarray}
where $N=3, n_f=5$.

Putting together all pieces of the real correction, we can see that
the IR divergences from the real correction can be written as

\begin{eqnarray}\label{realIR}
 \sigma^{Real}_{IR}&=&\frac{\alpha _s}{6 \pi }D_{\epsilon}
\Big\{\frac{13}{\epsilon _{\text{IR}}^2}+\left(4 \text{ln}(\frac{s}{m_t^2})
+\text{ln}(\frac{m_t^2-u}{m_t^2})-9 \text{ln}(
\frac{m_t^2-t}{m_t^2})+\frac{43}{2}\right)\frac{1}{\epsilon
_{\text{IR}}}\Big\}\sigma^B,
\end{eqnarray}
and all the IR divergences from the virtual corrections
in Eq.~(\ref{eq2}) are canceled exactly, as we expected.

\section{Contributions from the electroweak and strong FCNC couplings and the mixing effects}\label{s4}

In previous sections, we only consider the
contributions from the electroweak FCNC couplings,
$\kappa^{Z}_{tq}$. However, for the $tZ$ associated
production process, there are
additional contributions from the strong FCNC couplings,
$\kappa^{g}_{tq}$, and the mixing effects between these two
couplings. Since the magnitudes of the coefficients $\kappa^V_{tq}(V=Z,g)$ depend on
the underlying new physics, the mixing effects may be significant in certain model.
The $O(\alpha_s)$ corrections to the process $pp\longrightarrow tZ$ induced by $tqg$
are similar to ones induced by $tqZ$, so we don't show its analytical results, and only
present the mixing effects in this section.

\begin{figure}[h]
      \begin{center}
     \scalebox{0.5}{\includegraphics*{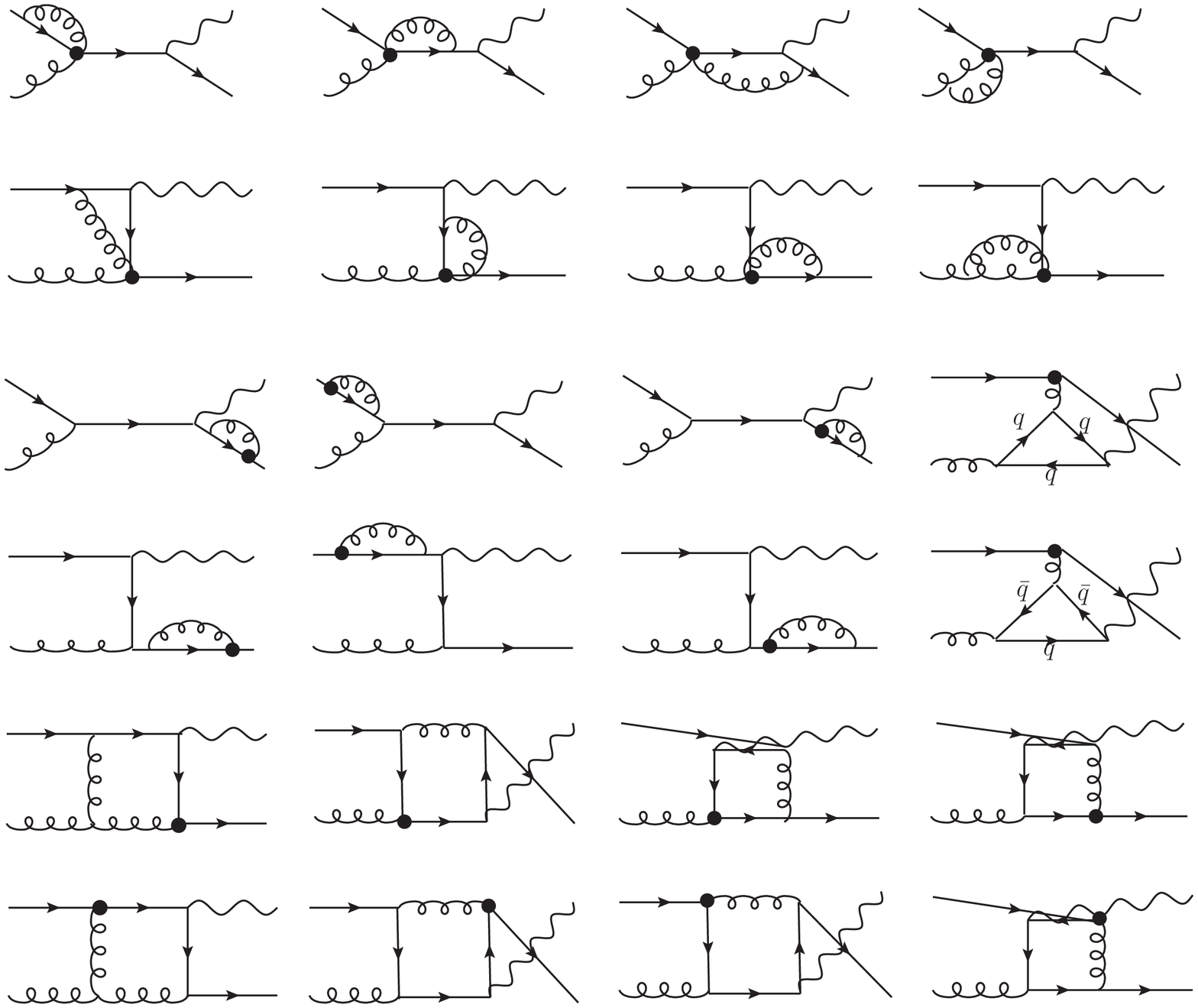}}
      \end{center}
    \caption[]{\label{finalvirtual}1-loop Feynman diagrams for the single top
quark production via the FCNC couplings with operator mixing.}
\end{figure}

\begin{figure}[h]
      \begin{center}
     \scalebox{0.6}{\includegraphics*{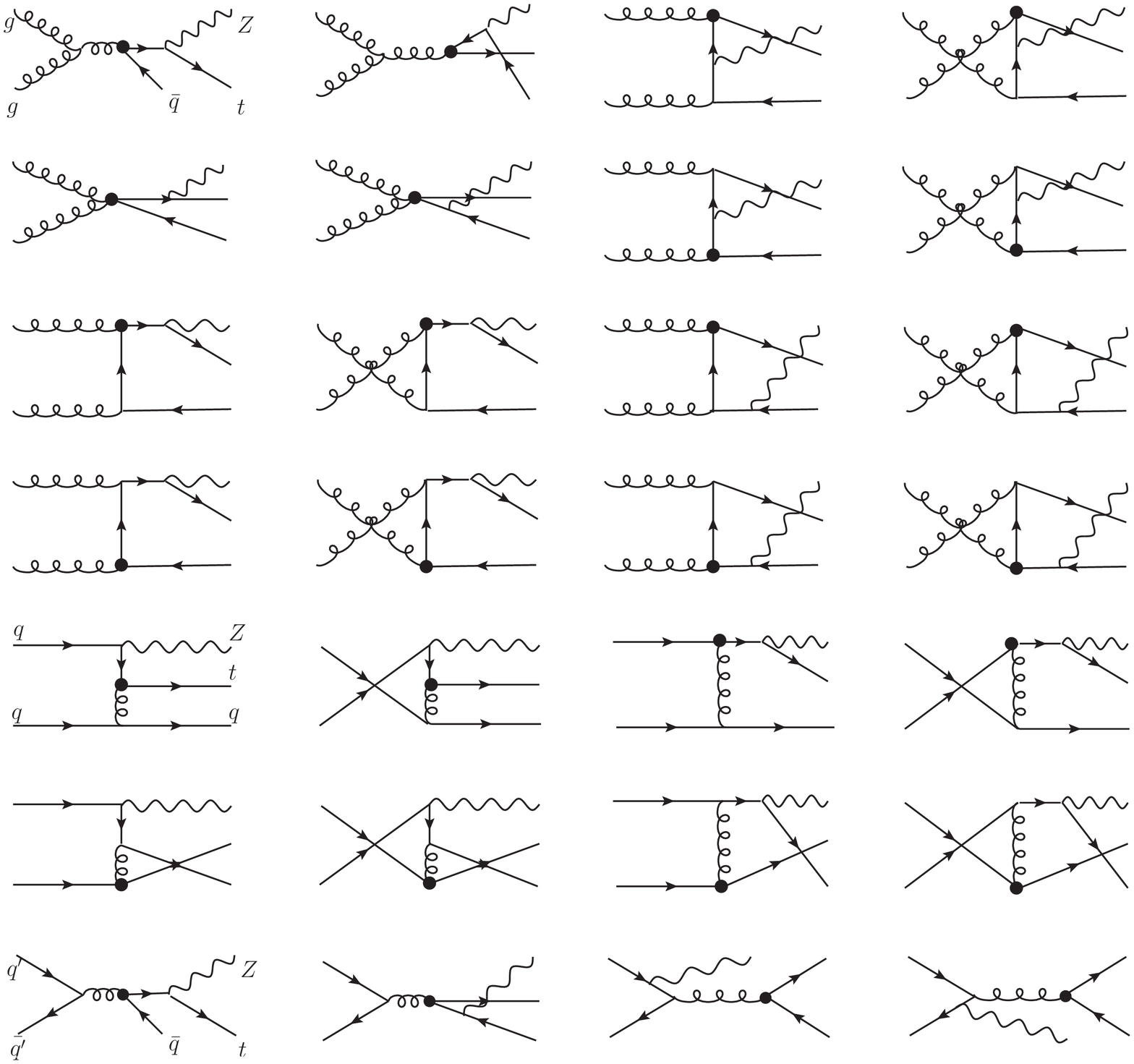}}
      \end{center}
    \caption[]{\label{finalrealhunhe}Feynman diagrams of the real corrections for the single top
quark production via the FCNC couplings with operator mixing.}
\end{figure}

At the LO, the contributing Feynman diagrams are show in Fig.~\ref{treehunhe},
and the squared amplitudes are present in the Appendix.
The NLO corrections, which
include the loop diagrams and the real emission diagrams, are shown in
Fig.~\ref{finalvirtual} and Fig.~\ref{finalrealhunhe}.

\begin{figure}[h]
\begin{center}
\scalebox{0.6}{\includegraphics*{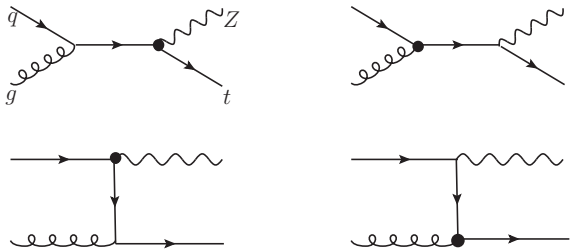}}
\caption[]{\label{treehunhe}The LO Feynman diagrams for the single top
quark production via the FCNC couplings with operator mixing.}
\end{center}
\end{figure}

The relevant renormalization constants are the same as ones in
Eq.~(\ref{deltaZ}) and~(\ref{deltaKz}), except
that we introduce additional renormalization constants. We adopt the
definition in Ref.~\cite{Zhang:2010bm}
\begin{equation}
\mathcal{L}_{\mathrm{eff}}+\delta\mathcal{L}_{\mathrm{eff}} = \left(
\begin{array}{ll}
 -\kappa ^g & -\kappa ^Z
\end{array}
\right)
\left(
\begin{array}{ll}
 1+\text{$\delta $Z}_{\text{gg}} & \text{$\delta
   $Z}_{\text{gZ}} \\
 \text{$\delta $Z}_{\text{Zg}} & 1+\text{$\delta
   $Z}_{\text{ZZ}}
\end{array}
\right)
\left(
\begin{array}{ll}
 O_g \\ O_Z
\end{array} \right),
\end{equation}
where the operators $O_i\ (i=g,Z)$ are defined as
$O_g=g_s\bar{q}\sigma^{\mu\nu}
T^a(f^g_{\mathrm{tq}}+ih^g_{\mathrm{tq}}\gamma_5)tG^a_{ \mu\nu}$,
$O_Z=\frac{e}{\text{sin2$\theta $}_W}
\bar{q}\sigma^{\mu\nu}(f^{Z}_{\mathrm{tq}}+ih^Z_{
\mathrm{tq}}\gamma_5) tZ_{\mu\nu}$, and $\delta Z_{gg}=\delta
Z_{\kappa^g_{\mathrm{tq}}/\Lambda}$, $\delta Z_{ZZ}=\delta
Z_{\kappa^Z_{\mathrm{tq}}/\Lambda}$. At the $\mathcal
{O}(\alpha_s)$ level, $\delta
Z_{\kappa^Z_{\mathrm{tq}}/\Lambda}$ is presented in
Eq.~(\ref{deltaKz}), and other renormalization constants are given
by:

\begin{eqnarray}
\delta
Z_{\kappa^{g}_{tq}/\Lambda}&=&\frac{\alpha_s}{6\pi}
\Gamma(1+\epsilon)(4\pi)
^{\epsilon}\frac{1}{\epsilon_{UV}}, \nonumber\\
\delta
Z_{gZ}&=&\frac{\alpha_s}{3\pi}\Gamma(1+\epsilon)(4\pi)
^{\epsilon}\frac{c_1+c_2}{\epsilon_{UV}}, \nonumber\\
 \delta Z_{Z g}&=&0,
\end{eqnarray}
$\delta Z_{gZ}$ is defined at the level of the cross section, and $c_1$, $c_2$ are defined as follows:
\begin{eqnarray}
c_1&=&g_{ZL}^*g_{gL}^{}Q_f sin\theta_W -g_{ZR}^*g_{gR}^{}\frac{s_3-Q_f sin^2\theta_W}{sin\theta_W}, \nonumber \\
c_2&=&g_{ZR}^*g_{gR}^{}Q_f sin\theta_W-g_{ZL}^*g_{gL}^{}\frac{s_3-Q_f sin^2\theta_W}{sin\theta_W}, \nonumber
\end{eqnarray}
where $Q_f$ is the electric charge of the fermion, and $s_3$ is its third component of the $SU(2)_L$ gauge group.

The renormalization group running for $\kappa^V$ are modified to~\cite{Zhang:2010bm}:
\begin{eqnarray}
\frac{\kappa^g_{tq}(\mu)}{\Lambda} &=& \frac{\kappa^g_{tq}(\mu')}{\Lambda}
\eta^{\frac{2}{3\beta_0}}, \nonumber \\
\frac{\kappa^Z_{tq}(\mu)}{\Lambda} &=& \frac{\kappa^Z_{tq}(\mu')}{\Lambda}
\eta^{\frac{4}{3\beta_0}}+
\frac{\kappa^g_{tq}(\mu')}{\Lambda}(\frac{32}{3}sin^2\theta_W-4)
 \left(\eta^\frac{4}{3\beta_0}
-\eta^{\frac{2}{3\beta_0}}\right),
\end{eqnarray}
where $\eta=\frac{\alpha_s(\mu')}{\alpha_s(\mu)}$.

The IR divergence appearing in the virtual corrections has the same form
as given in section~\ref{s3}, except using the LO
amplitude including the contributions from both the electroweak and
the strong FCNC couplings instead of $\hat\sigma^B$.

\section{Numerical Results}\label{s5}

\subsection{Process via $tqZ$  FCNC
couplings  without operator mixing effects}
We first consider the $tZ$ associated production via the $tqZ$ FCNC
couplings, including the NLO QCD effects on the
 total cross sections, the scale dependence, and several
important distributions at both the Tevatron and LHC.
All the SM input parameters are taken
to be~\cite{Nakamura:2010zzi}:
\begin{equation}
m_t=172.0{\rm GeV},\quad \alpha_s(M_Z)=0.118,\quad \alpha=1/128.921.
\end{equation}
And we set the electroweak FCNC couplings, allowed by current experiment, as follows:
\begin{equation}
\kappa^Z_{tu}/\Lambda=\kappa^Z_{tc}/\Lambda=0.5{\rm TeV}^{-1}.
\end{equation}
The running QCD coupling constant is evaluated at the three-loop
order~\cite{Amsler:2008zzb} and the CTEQ6M PDF
set~\cite{Pumplin:2002vw} is used throughout the calculations of the
NLO (LO) cross sections. Both the renormalization and factorization
scales are fixed to the sum of the top quark and the $Z$ boson mass.

In Table~\ref{table1}, we list some typical numerical results of the LO
and NLO total cross sections for the $tZ$ associated production via
the electroweak FCNC couplings. It can be seen that, for the
$tuZ$ coupling the NLO corrections can enhance the total cross
sections by about 60\% and 42\%, and for the $tcZ$ coupling by about
51\% and 43\% at the Tevatron and LHC, respectively.

\begin{table}[h]
\begin{center}
\begin{tabular}{ccccc}
  \hline
  \hline
  FCNC coupling & $tuZ$  (LO) & $tuZ$  (NLO) & $tcZ$  (LO) & $tcZ$  (NLO) \\
  \hline
  LHC  (pb) & 15.9 & 22.5 & 1.29 & 1.85 \\
  \hline
  Tevatron  (fb) & 55.5  & 88.6  & 1.62 & 2.45  \\
  \hline
  \hline
\end{tabular}
\end{center}
\caption{The LO and NLO total cross sections for $tZ$
associated production via the electroweak FCNC couplings at both the LHC and
Tevatron. Here
$\mu=m_t+m_z$, $\kappa^Z_{\mathrm{tq}}/\Lambda=
\mathrm{0.5TeV}^{-1}$.} \label{table1}
\end{table}

\begin{figure}[h]
      \begin{center}
     \scalebox{0.32}{\includegraphics*{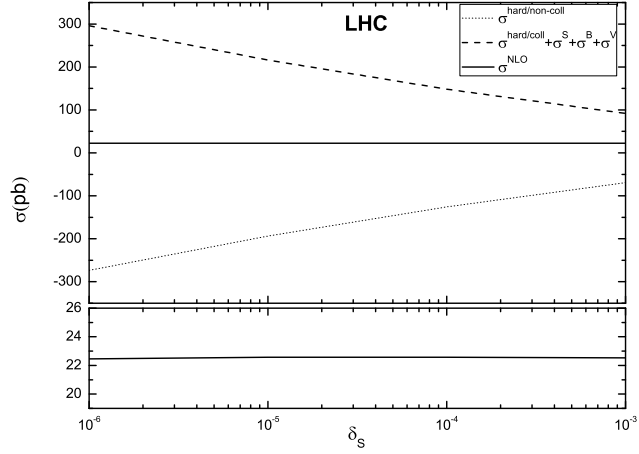}}
      \end{center}
    \caption{\label{udeltas}Inclusive total cross sections for $pp\rightarrow tZ+X$ at the
     LHC as a function of $\delta_{s}$ in the phase space slicing treatment.
  The $\delta_{c}$  is chosen to be $\delta_{c}=\delta_{s}/50$.}
\end{figure}

In Fig.\ref{udeltas} we show that it is reasonable to use the two
cutoff phase space slicing method in our NLO QCD calculations; i.e.,
the dependence of the NLO QCD predictions on the arbitrary cutoffs
$\delta_s$ and $\delta_c$ is indeed very weak. While the Born cross sections and the
virtual corrections are cutoff independent, both the soft and
collinear contributions and the noncollinear contributions depend
strongly on the cutoffs. However, the cutoff dependence in the two
contributions ($\sigma^{S} + \sigma^{coll}$ and
$\sigma^{\overline{HC}} + \sigma^{\overline{C}}$) nearly cancel
each other, and the final results for $\sigma^{NLO}$ are almost
independent of the cutoffs. We will take
$\delta_s = 10^{-4}$ in the numerical calculations below. Generally
$\delta_c$ being $50-100$ times smaller than $\delta_s$ is
sufficient for accurate calculations to a few
percent\cite{Harris:2001sx}, so we take $\delta_c = \delta_s/50$ in
our calculations.

\begin{figure}[H]
  \subfigure{
    \begin{minipage}[b]{0.5\textwidth}
      \begin{center}
     \scalebox{0.3}{\includegraphics*{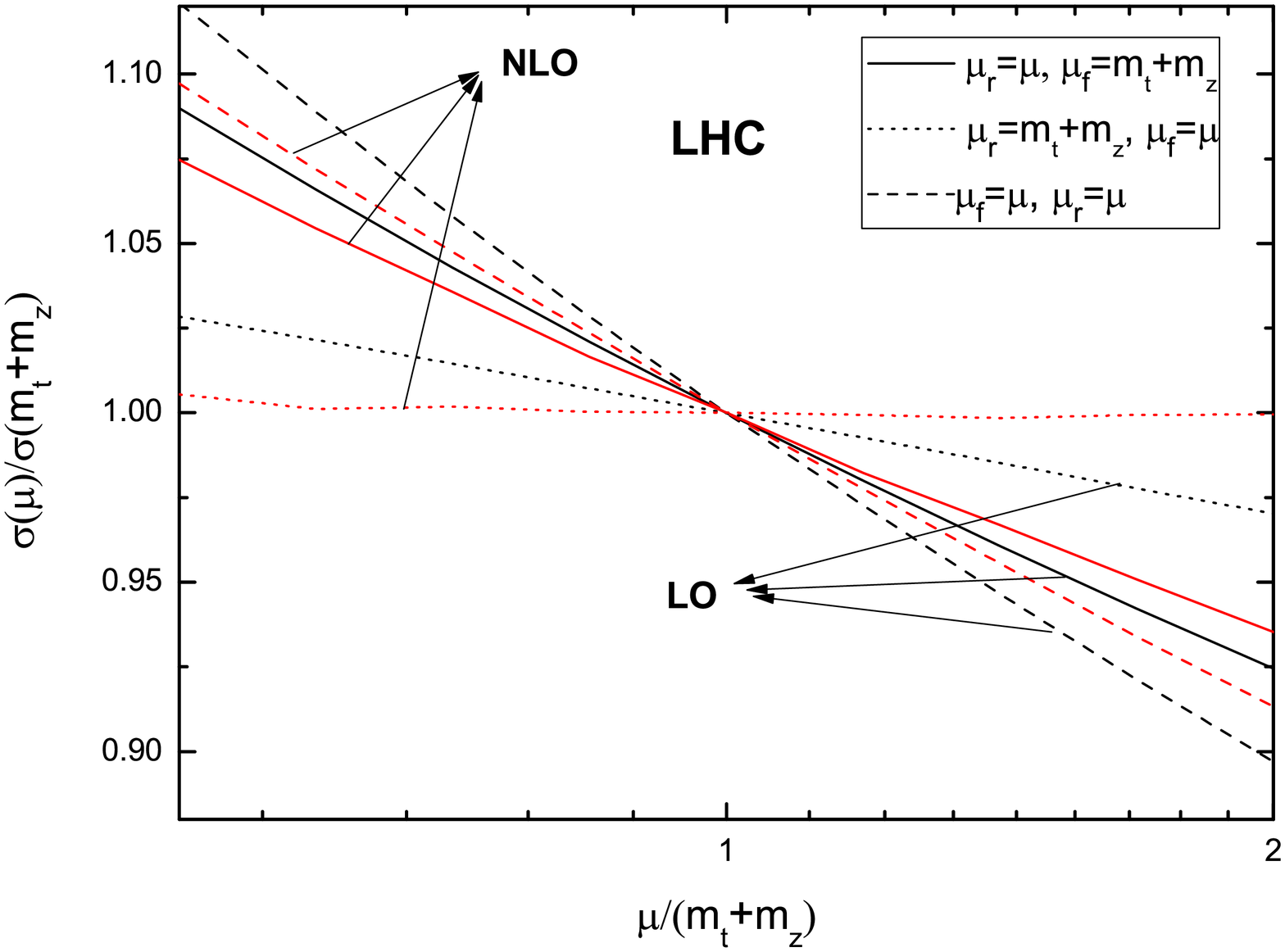}}
      \end{center}
    \end{minipage}}
  \subfigure{
    \begin{minipage}[b]{0.5\textwidth}
      \begin{center}
     \scalebox{0.3}{\includegraphics*{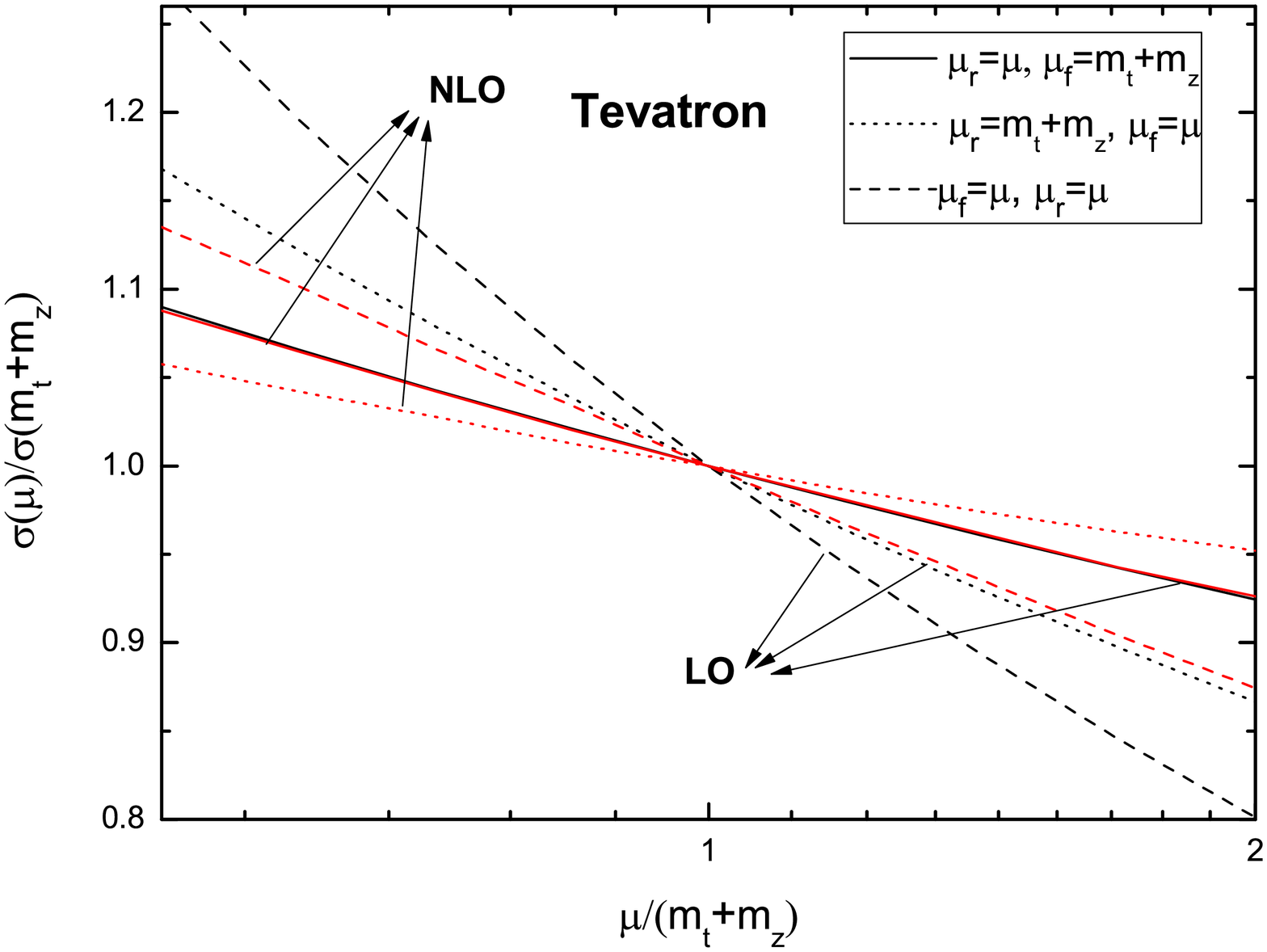}}
      \end{center}
    \end{minipage}}
    \caption{\label{scalez}Scale dependence of the total cross sections for the
     $gu$ initial state subprocess at both LHC and Tevatron,
    the black lines represent the LO results, while the red ones represent
    the NLO results.}
\end{figure}

\begin{figure}[H]
  \subfigure{
    \begin{minipage}[b]{0.5\textwidth}
      \begin{center}
     \scalebox{0.28}{\includegraphics*{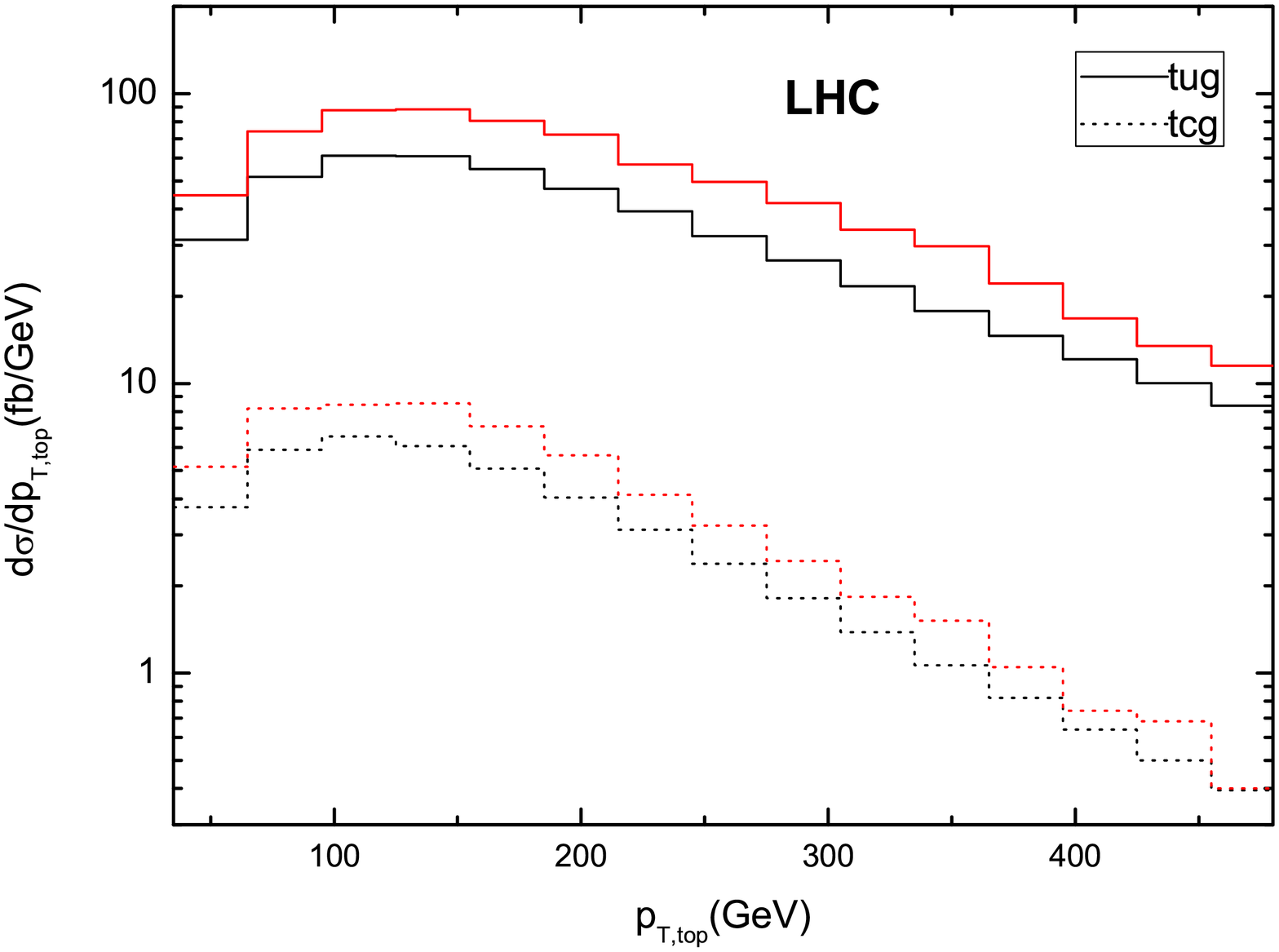}}
      \end{center}
    \end{minipage}}
  \subfigure{
    \begin{minipage}[b]{0.5\textwidth}
      \begin{center}
     \scalebox{0.28}{\includegraphics*{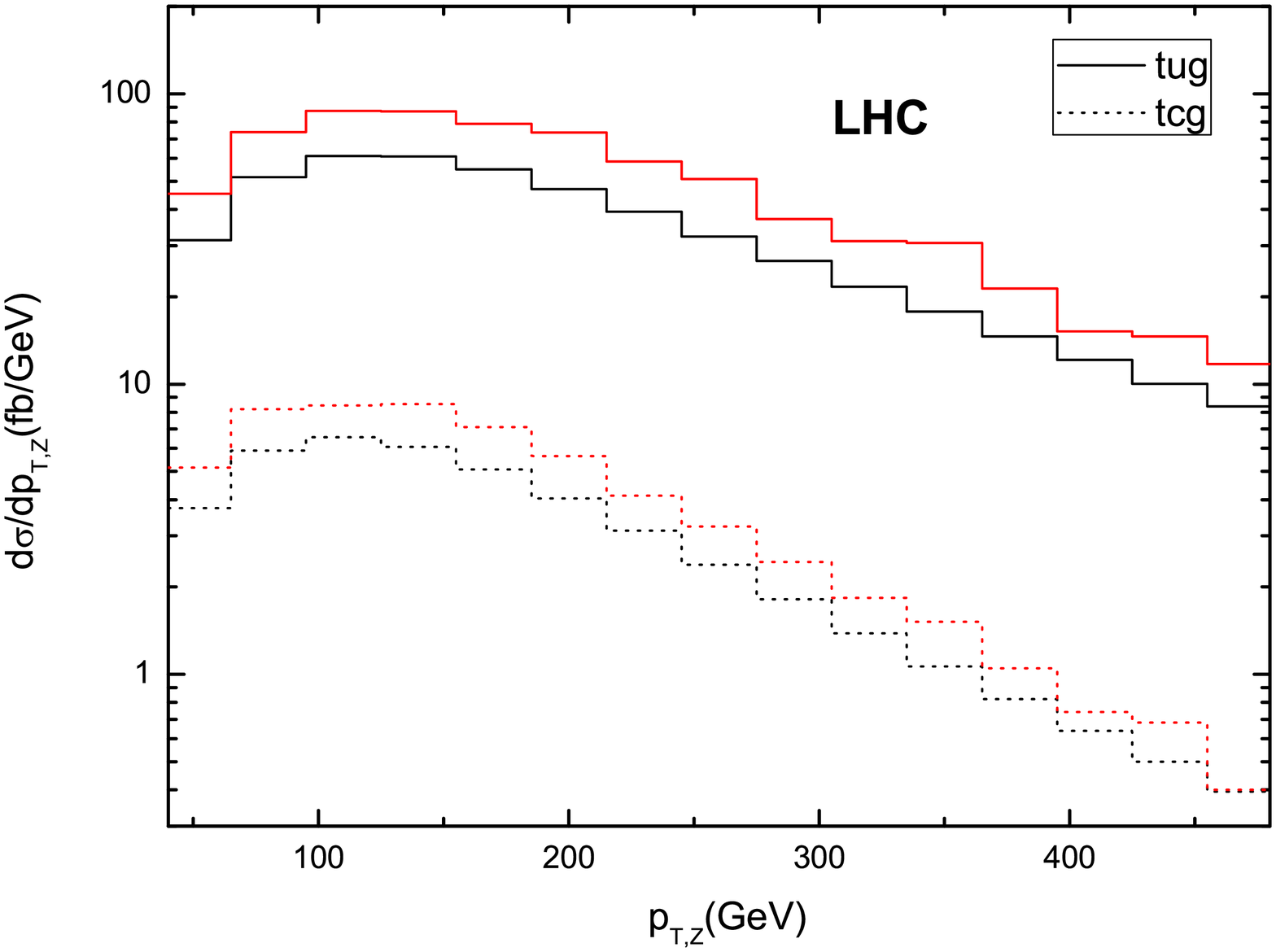}}
      \end{center}
    \end{minipage}}
    \caption{\label{f4}$p_T$ distributions of the top
    quark and the Z boson at LHC, the black and red lines represent the LO and the NLO results of the FCNC
    single top production, respectively.}
\end{figure}

\begin{figure}[H]
      \begin{center}
     \scalebox{0.32}{\includegraphics*[30,15][920,580]{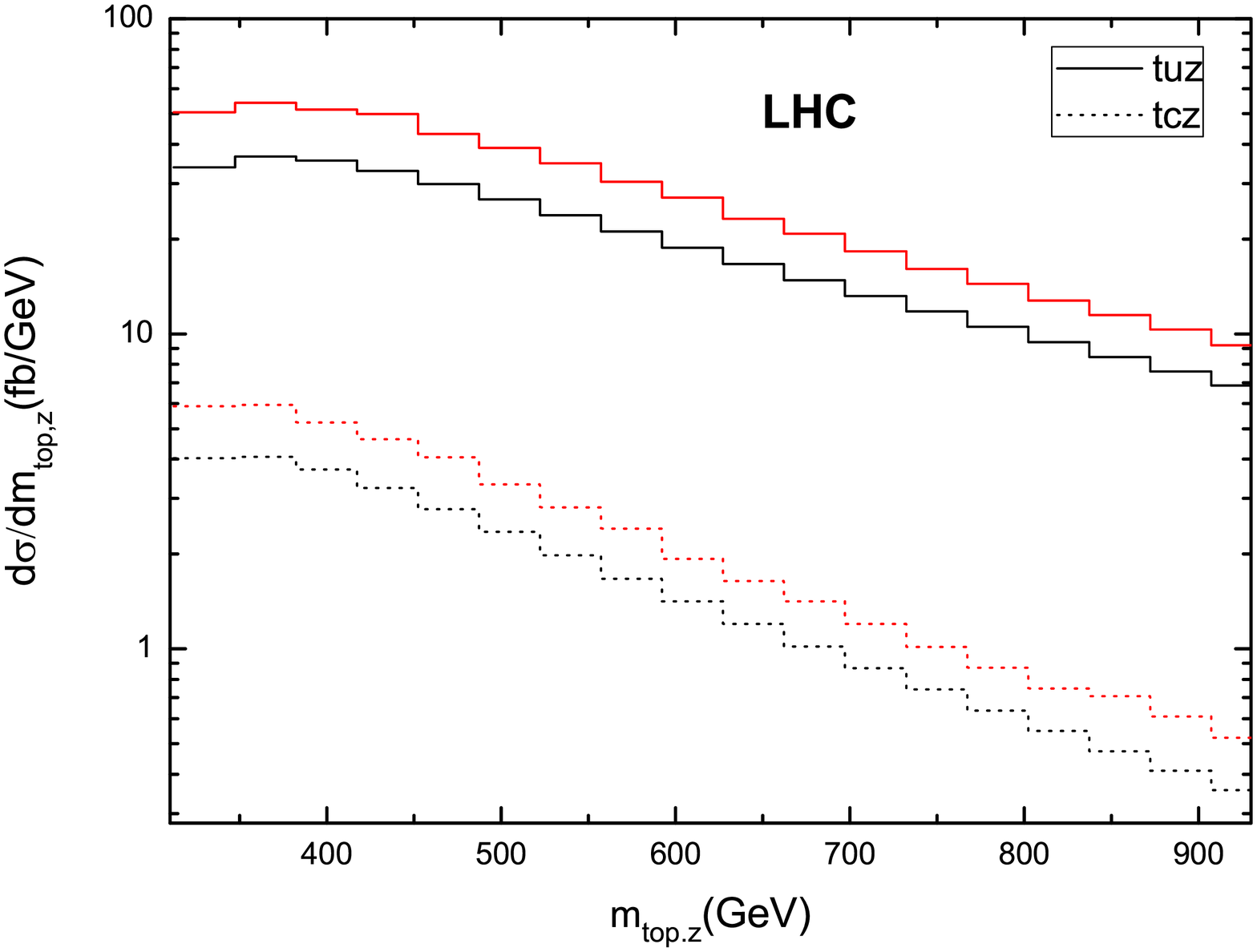}}
      \end{center}
    \caption{\label{f5}Invariant mass  distributions of the Z
    boson and the top quark, the black and red lines represent the LO and the NLO results of the FCNC
    single top production, respectively.}
\end{figure}

In Figs.~\ref{scalez} we show the scale dependence of the
LO and NLO total cross section for three cases: (1) the
renormalization scale dependence $\mu_r=\mu,\ \mu_f=m_t+m_Z$; (2) the
factorization scale dependence $\mu_r=m_t+m_Z,\ \mu_f=\mu$; and (3) total
scale dependence $\mu_r=\mu_f=\mu$. It can be seen that the NLO
corrections reduce the scale dependence for all three
cases, which make the theoretical predictions more reliable.
Fig.~\ref{f4} give the $p_T$ distributions of the top quark and the $Z$ boson, respectively,
and Fig~\ref{f5} shows the invariant mass distributions of the $Z$ boson and the top quark.

\subsection{Process include operator mixing effects}
In this subsection, we present the numerical results of the $tZ$
associated production via the electroweak and strong FCNC couplings,
including the NLO QCD effects and the mixing effects.
We investigate the NLO QCD effects on the total cross sections
and the scale dependence at the LHC. For the numerical calculations,
we take the same SM parameters as above subsection.

In Table~\ref{table2}, we list some typical numerical results of the LO
and NLO total cross sections for the $tZ$ associated production via
the strong FCNC couplings, assuming $\kappa^g_{\mathrm{tq}}/\Lambda=\mathrm{0.01TeV}^{-1}$,
allowed by current experiment. It can be seen that, the NLO corrections
can enhance the total cross sections by about 27\% for the $tug$ couplings,
and by about 42\% for the $tcg$ coupling at the LHC. Nevertheless,
the contributions to the total cross sections from the process induced by the $tqZ$ FCNC couplings
are still dominate.

\begin{table}[h]
\begin{center}
\begin{tabular}{ccccc}
\hline
  \hline
  FCNC coupling & $tug$  (LO) & $tug$  (NLO) & $tcg$  (LO) & $tcg$  (NLO) \\
  \hline
  LHC  (fb) & 141 & 180 & 7.6 & 10.8 \\
  \hline
  \hline
\end{tabular}
\end{center}
\caption{The LO and the NLO total cross sections for the $tZ$
associated production via the strong FCNC couplings at the LHC. Here
$\mu=m_t+m_z$, $\kappa^g_{\mathrm{tq}}/\Lambda=
\mathrm{0.01TeV}^{-1}$, and $|g_{gL}|^2=|g_{gR}|^2=1$.} \label{table2}
\end{table}

After considering the mixing effects, the total cross sections of
the $tZ$ associated production via FCNC couplings
can be factorized as:

\begin{eqnarray}
\sigma&=& A_1 |g_{\text{gL}}|^2 (\frac{\kappa^{g}_{tq}}{\text{$\Lambda $}})^2
+ A_2 |g_{\text{gR}}|^2(\frac{ \kappa^{g}_{tq}}{\text{$\Lambda $}})^2
+ A_3 (\frac{ \kappa^{Z}_{tq}}{\Lambda})^2 + \nonumber \\ &&
 \big[A_4 Re(g_{\text{gL}}g_{\text{ZL}}^*)
- A_5 Re(g_{\text{gR}} g_{\text{ZR}}^*)\big]\frac{ \kappa^{g}_{tq} \kappa^{Z}_{tq}}{\text{$\Lambda $}^2},
\end{eqnarray}
where $A_i$ represent the contributions from different
couplings and mixing effects. And, their numerical expressions at
the LHC can be written as
\begin{eqnarray}
\sigma^{tuV}_{LO} &=& \Big\{
715 |g_{\text{gL}}|^2 (\frac{\kappa^{g}_{tu}}{\text{$\Lambda $}})^2
+ 699 |g_{\text{gR}}|^2(\frac{ \kappa^{g}_{tu}}{\text{$\Lambda $}})^2
+ 63.7 (\frac{ \kappa^{Z}_{tu}}{\Lambda})^2 \nonumber \\ &&
+ \big[76.0 Re(g_{\text{gL}}g_{\text{ZL}}^*)
- 84.1 Re(g_{\text{gR}} g_{\text{ZR}}^*)\big]\frac{ \kappa^{g}_{tu} \kappa^{Z}_{tu}}{\text{$\Lambda $}^2}
\Big\} (\text{pb} \cdot \rm{TeV^2}), \nonumber \\ \\
\sigma^{tuV}_{NLO} &=& \Big\{
925 |g_{\text{gL}}|^2 (\frac{\kappa^{g}_{tu}}{\text{$\Lambda $}})^2
+ 874 |g_{\text{gR}}|^2(\frac{ \kappa^{g}_{tu}}{\text{$\Lambda $}})^2
+ 90 (\frac{ \kappa^{Z}_{tu}}{\Lambda})^2 \nonumber \\ &&
+ \big[107 Re(g_{\text{gL}}g_{\text{ZL}}^*)
- 114 Re(g_{\text{gR}} g_{\text{ZR}}^*)\big]\frac{ \kappa^{g}_{tu} \kappa^{Z}_{tu}}{\text{$\Lambda $}^2}
\Big\} (\text{pb} \cdot \rm{TeV}^2), \nonumber \\ \\
\sigma^{tcV}_{LO} &=& \Big\{
38.9 |g_{\text{gL}}|^2 (\frac{\kappa^{g}_{tc}}{\text{$\Lambda $}})^2
+ 37.2 |g_{\text{gR}}|^2(\frac{ \kappa^{g}_{tc}}{\text{$\Lambda $}})^2
+ 5.15 (\frac{ \kappa^{Z}_{tc}}{\Lambda})^2 \nonumber \\ &&
+ \big[6.97 Re(g_{\text{gL}}g_{\text{ZL}}^*)
- 7.47 Re(g_{\text{gR}} g_{\text{ZR}}^*)\big]\frac{ \kappa^{g}_{tc} \kappa^{Z}_{tc}}{\text{$\Lambda $}^2}
\Big\} (\text{pb} \cdot \rm{TeV^2}), \nonumber \\ \\
\sigma^{tcV}_{NLO} &=& \Big\{
56.7 |g_{\text{gL}}|^2 (\frac{\kappa^{g}_{tc}}{\text{$\Lambda $}})^2
+ 51.5 |g_{\text{gR}}|^2(\frac{ \kappa^{g}_{tc}}{\text{$\Lambda $}})^2
+ 7.38 (\frac{ \kappa^{Z}_{tc}}{\Lambda})^2 \nonumber \\ &&
+ \big[9.13 Re(g_{\text{gL}}g_{\text{ZL}}^*)
- 9.83 Re(g_{\text{gR}} g_{\text{ZR}}^*)\big]\frac{ \kappa^{g}_{tc} \kappa^{Z}_{tc}}{\text{$\Lambda $}^2}
\Big\} (\text{pb} \cdot \rm{TeV^2}), \nonumber \\
\end{eqnarray}
where $g_{iL}, g_{iR}$ are chiral parameters:
$$
g_{iL}=f^i_{\mathrm{tq}}-ih^i_{\mathrm{tq}},\qquad
g_{iR}=f^i_{\mathrm{tq}}+ih^i_{\mathrm{tq}},\qquad
|g_{iL}|^2+|g_{iR}|^2=2.
$$

In Table~\ref{table3}, we list some typical numerical results of the LO
and NLO total cross sections by choosing a special set of parameters
(for simplicity, we set $\kappa^Z_{\mathrm{tq}}/\Lambda=\kappa^g_{\mathrm{tq}}/\Lambda=\mathrm{0.01TeV}^{-1}$) and
fix $\mu=m_t+m_Z$.
\begin{table}[t]
\begin{center}
\begin{tabular}{ccccc}
  \hline
  \hline
  FCNC coupling & $tuV$  (LO) & $tuV$  (NLO) & $tcV$  (LO) & $tcV$  (NLO) \\
  \hline
  LHC  (fb) &
   147 & 188 & 8.1 & 11.5 \\
  \hline
  \hline
\end{tabular}
\end{center}
\caption{The LO and the NLO total cross sections for the $tZ$
associated production via the FCNC couplings at the LHC. Here
$\mu=m_t+m_z$, $\kappa^V_{\mathrm{tq}}/\Lambda=
\mathrm{0.01TeV}^{-1}$ and
$|g_{gL}|^2=|g_{gR}|^2=Re(g_{gL}g_{ZL}^*)=Re(g_{gR}g_{ZR}^*)=1$ \label{table3}.}
\end{table}
For the $g \ u \rightarrow t \ Z$, the NLO corrections
can enhance the total cross sections by about 28\%,
and for the $g \ c \rightarrow t \ Z$ process, by about
42\% at the LHC.

\begin{figure}[H]
  \subfigure{
    \begin{minipage}[b]{0.5\textwidth}
      \begin{center}
     \scalebox{0.5}{\includegraphics*{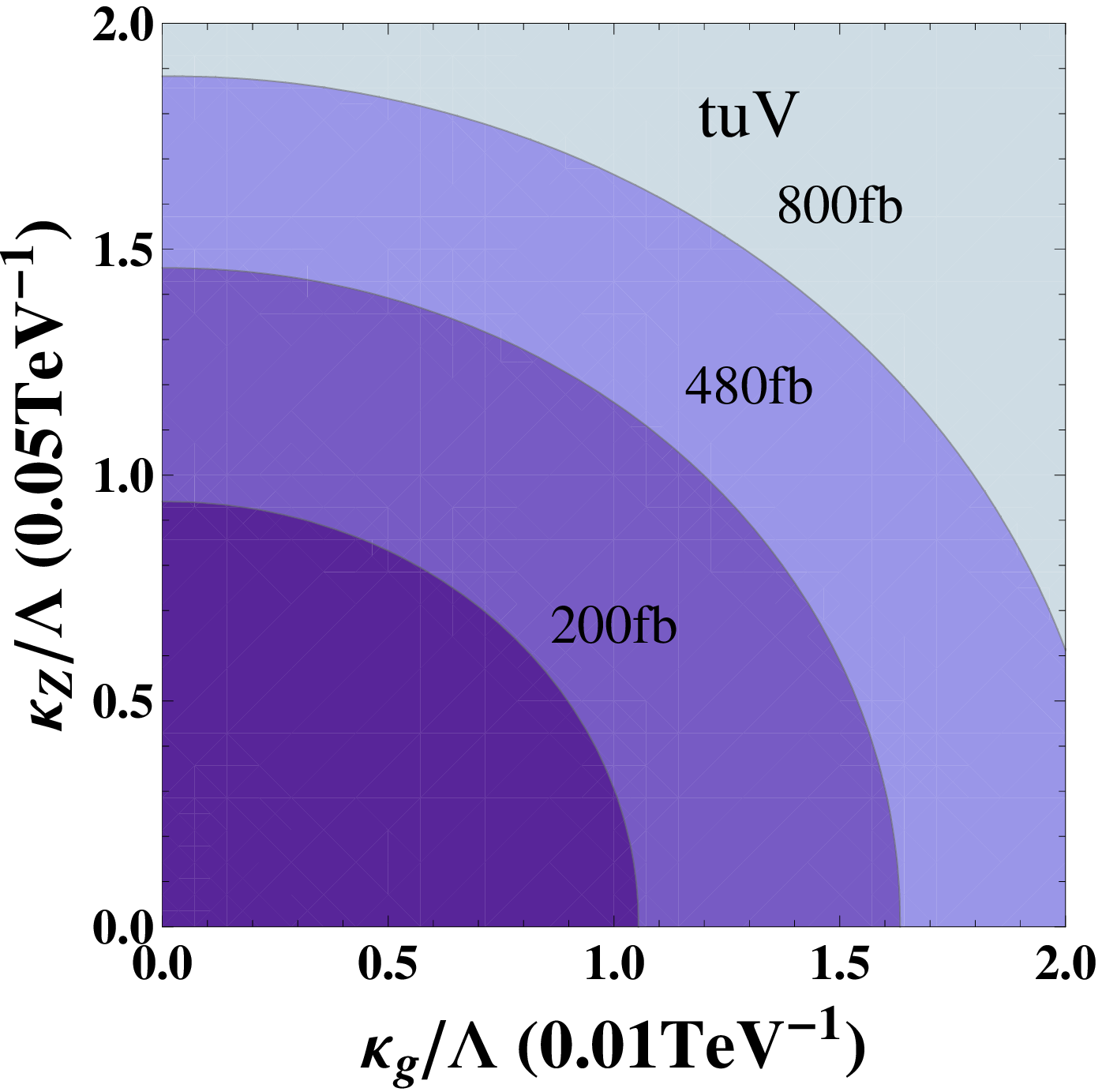}}
      \end{center}
    \end{minipage}}
  \subfigure{
    \begin{minipage}[b]{0.5\textwidth}
      \begin{center}
     \scalebox{0.5}{\includegraphics*{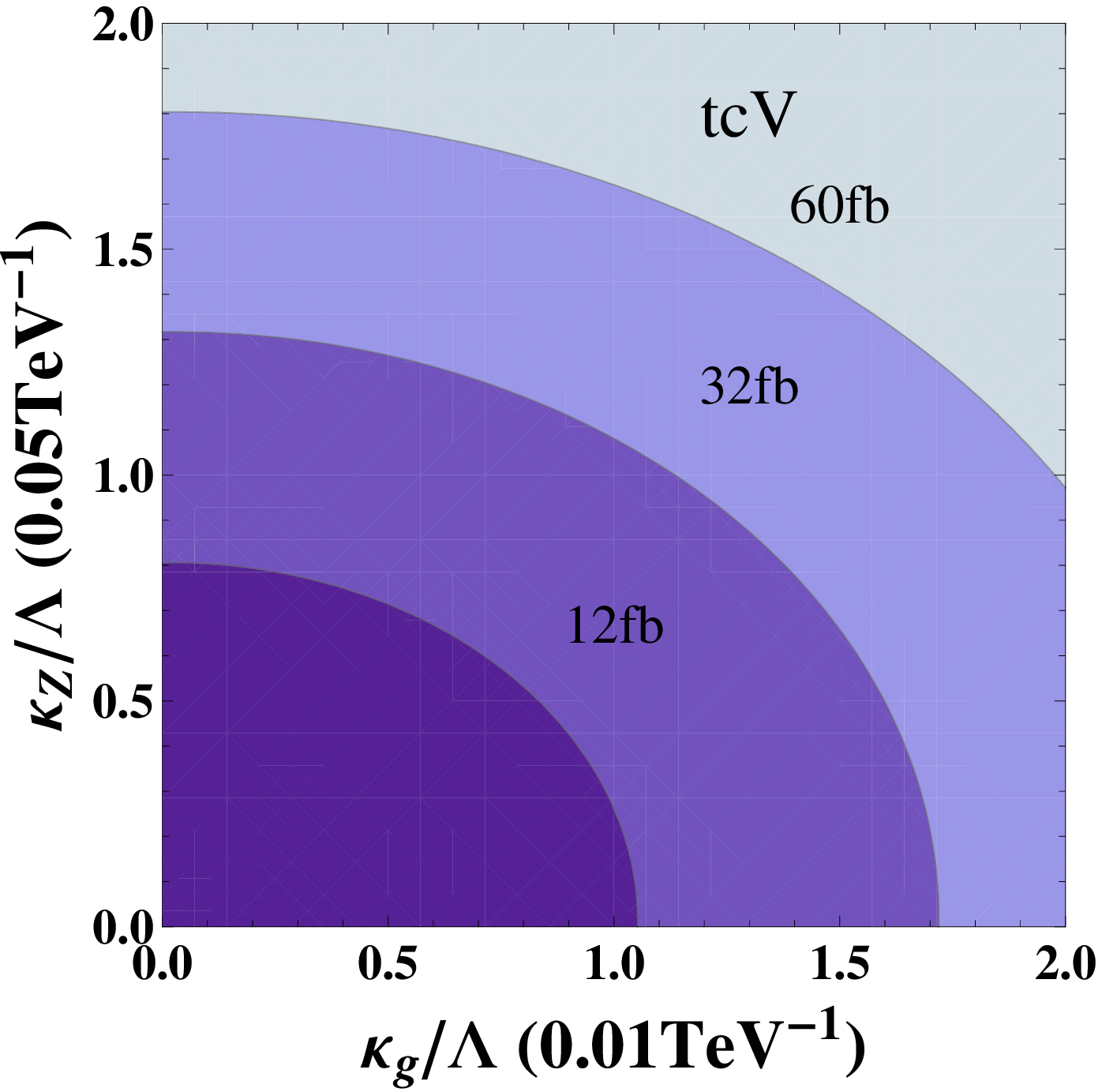}}
      \end{center}
    \end{minipage}}
    \caption{\label{uKgKz}The contour curves of the total cross sections versus
    the parameters $\kappa_g/\Lambda$ and $\kappa_z/\Lambda$ for the process
    induced by tuV and tcV FCNC couplings. Here we set $|g_{gL}|^2=|g_{gR}|^2=Re(g_{gL}g_{ZL}^*)=Re(g_{gR}g_{ZR}^*)=1$.}
\end{figure}

\begin{figure}[H]
  \subfigure{
    \begin{minipage}[b]{0.5\textwidth}
      \begin{center}
     \scalebox{0.5}{\includegraphics*{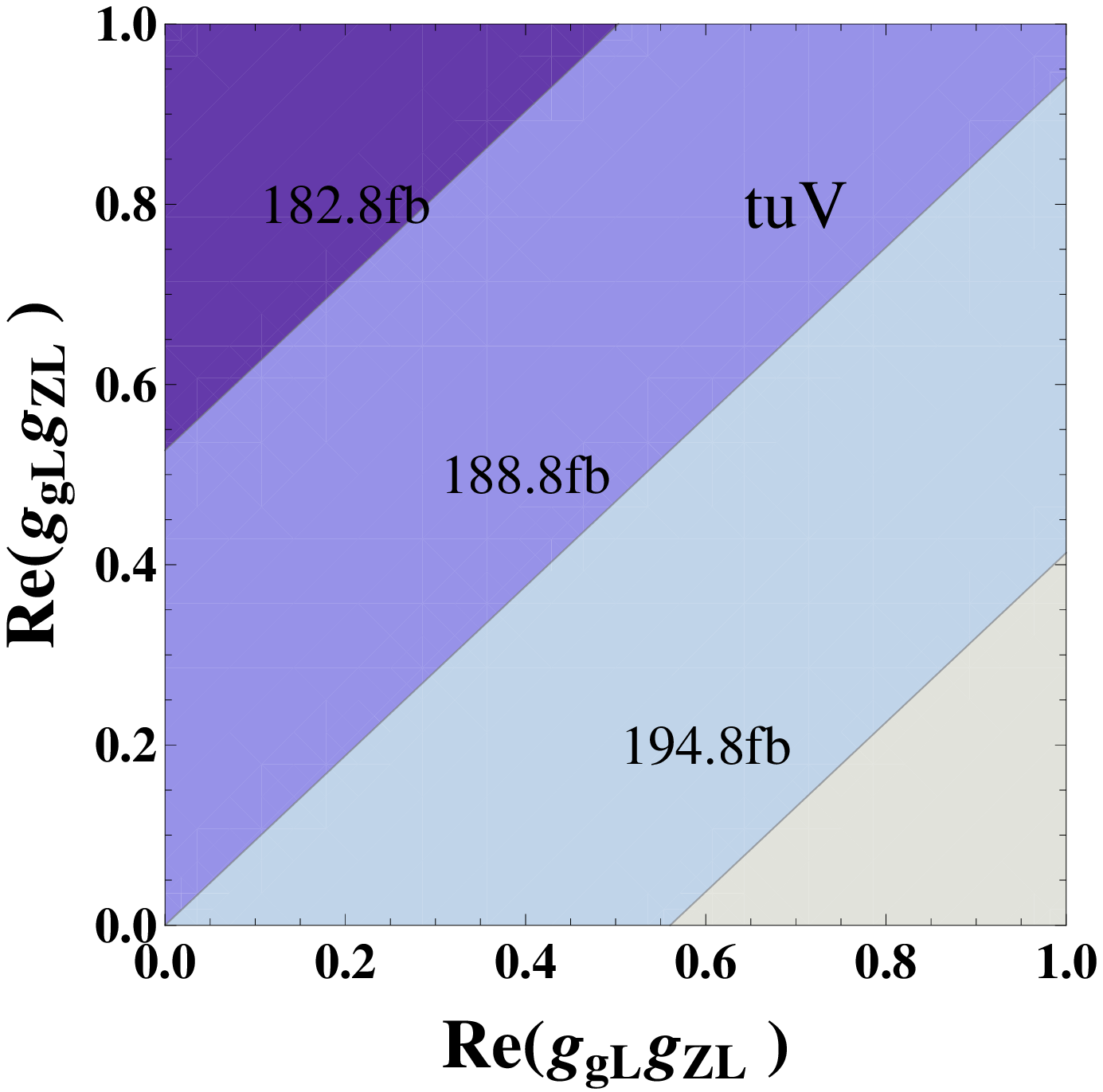}}
      \end{center}
    \end{minipage}}
  \subfigure{
    \begin{minipage}[b]{0.5\textwidth}
      \begin{center}
     \scalebox{0.5}{\includegraphics*{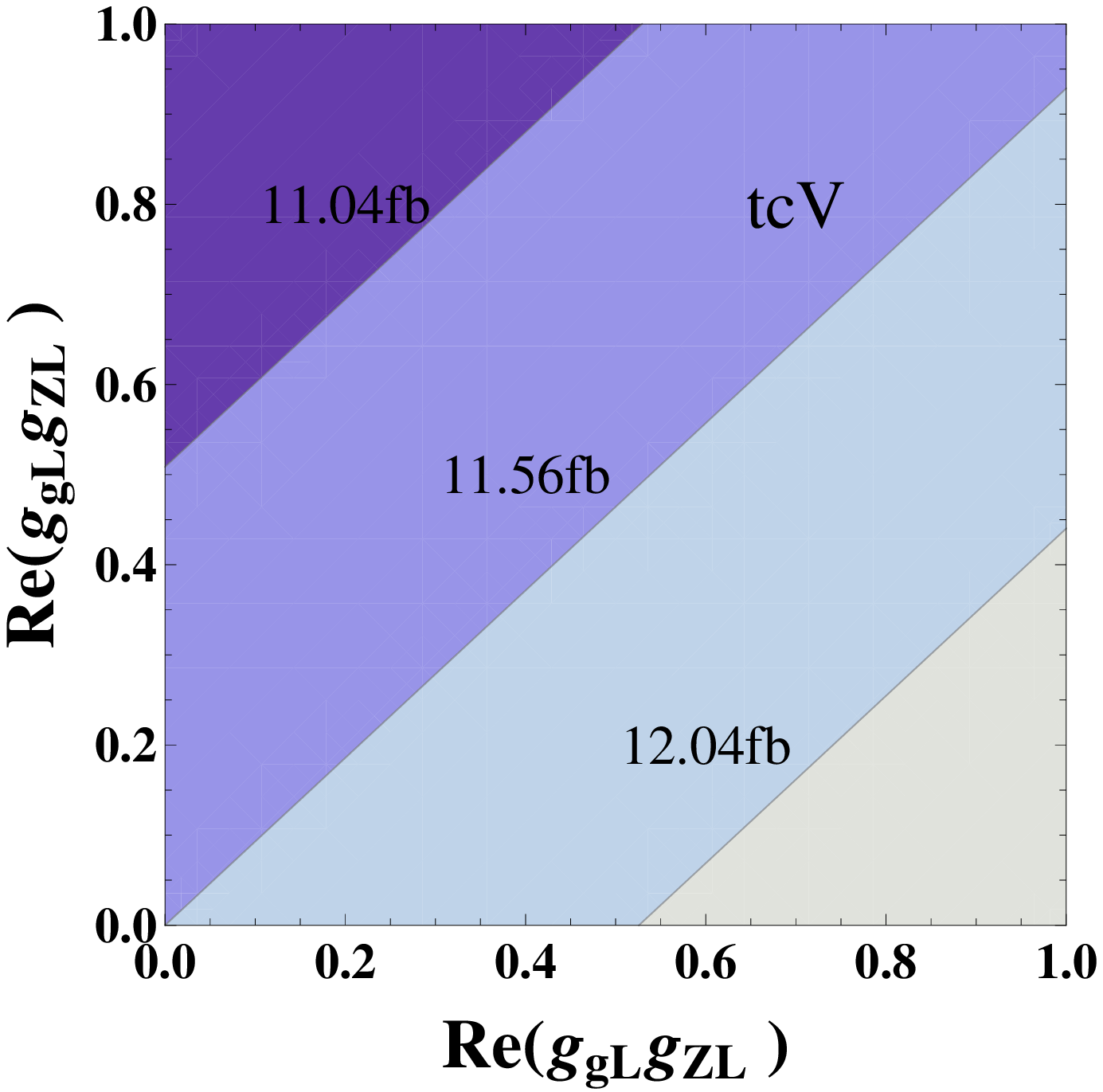}}
      \end{center}
    \end{minipage}}
    \caption{\label{ulr}The contour curves of the total cross sections versus
    the parameters $Re(g_{gL}\times g_{ZL})$ and $Re(g_{gR}\times g_{ZR})$ for the process
    induced by tuV and tcV FCNC couplings.
    Here we set $|g_{gL}|^2=|g_{gR}|^2=|g_{ZL}|^2=|g_{ZR}|^2=1$ and $\kappa_g/\Lambda=\kappa_z/\Lambda=
    0.01\rm{TeV^{-1}}$.}
\end{figure}

To investigate the contributions from the operator mixing effects, we present the counter curves
for the variables $\kappa_Z/\Lambda, \kappa_g/\Lambda$, and $Re(g_{gL}^*g_{ZL}),Re(g_{GR}^*g_{ZR})$ in Figs.~\ref{uKgKz},~\ref{ulr}

\begin{figure}[H]
      \begin{center}
     \scalebox{0.32}{\includegraphics*{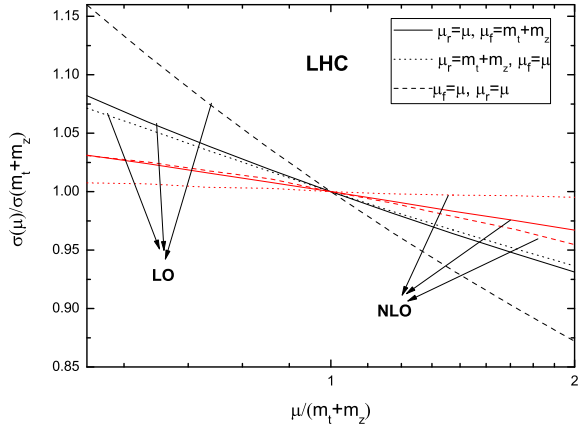}}
      \end{center}
    \caption{\label{f8}Scale dependence of the total cross sections for the
    $gu$ initial state subprocess at the LHC,
    the black lines represent the LO results, while the red ones represent
    the NLO results.}
\end{figure}

In Figs.~\ref{f8} we show the scale dependence of the
LO and NLO total cross section at LHC for three cases: (1) the
renormalization scale dependence $\mu_r=\mu,\ \mu_f=m_t+m_Z$; (2) the
factorization scale dependence $\mu_r=m_t+m_Z,\ \mu_f=\mu$; and (3) total
scale dependence $\mu_r=\mu_f=\mu$. It can be seen that the NLO
corrections reduce the scale dependence significantly for all three
cases, which make the theoretical predictions more reliable.

\section{Conclusions}\label{s6}

We have calculated the NLO QCD corrections to the $tZ$
associated production via the $tqZ$ and $tqg$ FCNC
couplings at hadron colliders, respectively, and we also consider
the mixing effects of these two couplings. Our results show that, for the $tuZ$ coupling
the NLO corrections can enhance the total cross sections by about
60\% and 42\%, and for the $tcZ$ coupling by about 51\% and 43\% at
the Tevatron and LHC, respectively. The NLO corrections can enhance the total cross sections by about 27\% and 42\%
for the $tug$ and the $tcg$ couplings, respectively, at the LHC.
The mixing effects between the $tqZ$ and $tqg$ FCNC couplings for this process
can be either large or small depending on the values of various
anomalous couplings. If we set $\kappa_g/\Lambda=\kappa_z/\Lambda=0.01\rm{TeV}^{-1}$ and
$|g_{gL}|^2=|g_{gR}|^2=Re(g_{gL}g_{ZL}^*)=Re(g_{gR}g_{ZR}^*)=1$, the
NLO corrections can enhance the total cross sections by
about 28\% for $tuV$ couplings, and by 42\% for $tcV$ couplings at the LHC.
Moreover, the NLO corrections reduce
the dependence of the total cross sections on the renormalization or
factorization scale significantly.

\begin{acknowledgments}

This work was supported in part by National Nature Science Foundation of China,
Under Grants No. 10975004, and No. 11021092.

\end{acknowledgments}

\section*{APPENDIX}
In this appendix we give the LO results of the $tZ$
associated production induced by the $tqg$ and $tqZ$ FCNC couplings, and
the definition of the A and $B_\pm$.

\begin{eqnarray}
\overline{|M^B|^2}_{gq}(s,t)&=&\frac{16 \pi ^2 \alpha  \alpha _s}{27 s t^2 \Lambda ^2 \sin (2 \theta )^2
   (s-m_t^2)^2 (t-m_t^2)^2 m_z^2}\{(9 t^2 g_{\text{ZL}}^2 (s-m_t^2)^2 (2 m_t^8-(3
\nonumber \\ &&   m_z^2+4 s+2 t) m_t^6+(2 m_z^4-(2 s+t) m_z^2+2 (s^2+4 t s+t^2)) m_t^4+(2
\nonumber \\ &&   m_z^6-4 t m_z^4+(s^2+6 t s+5 t^2) m_z^2-2 t (3 s^2+6 t s+t^2)) m_t^2+t
\nonumber \\ &&   (-2 m_z^6+2 (s+t) m_z^4-(s+t)^2 m_z^2+4 s t (s+t))) \kappa _z^2 m_z^2+9 t^2
\nonumber \\ &&   g_{\text{ZR}}^2 (s-m_t^2)^2 (2 m_t^8-(3 m_z^2+4 s+2 t) m_t^6+(2 m_z^4-(2
\nonumber \\ &&   s+t) m_z^2+2 (s^2+4 t s+t^2)) m_t^4+(2 m_z^6-4 t m_z^4+(s^2+6 t s+5 t^2)
\nonumber \\ &&   m_z^2-2 t (3 s^2+6 t s+t^2)) m_t^2+t (-2 m_z^6+2 (s+t) m_z^4-(s+t)^2 m_z^2+4 s t
\nonumber \\ &&   (s+t))) \kappa _z^2 m_z^2+6 s t g_{\text{gL}} g_{\text{ZL}} (s-m_t^2)
   (t-m_t^2) (-2 (4 \text{sw}^2-3) (3 t-m_z^2) m_t^6+2 (s
\nonumber \\ &&   (3-4 \text{sw}^2) m_z^2+t (s (16 \text{sw}^2-9)+4 (4 \text{sw}^2-3)
   t)) m_t^4+t (-8 s^2 \text{sw}^2+2 (3-4 \text{sw}^2)
\nonumber \\ &&    t^2+(-8 t\text{sw}^2+s (16 \text{sw}^2-9)+6 t) m_z^2+12 s (1-2 \text{sw}^2) t)m_t^2+s t^2 (8 s \text{sw}^2+
\nonumber \\ &&   8 t \text{sw}^2+(3-8 \text{sw}^2) m_z^2-6 t))
   \kappa _g \kappa _z m_z^2+6 s t g_{\text{gR}} g_{\text{ZR}} (s-m_t^2) (t-m_t^2)
\nonumber \\ &&   (8 \text{sw}^2 (m_z^2-3 t) m_t^6+(2 t (16 t \text{sw}^2+s (16
   \text{sw}^2-3))-8 s \text{sw}^2 m_z^2) m_t^4-t
\nonumber \\ &&   ((8 \text{sw}^2-6) s^2+6(4 \text{sw}^2-1) t s+8 \text{sw}^2 t^2+(8 t \text{sw}^2+s (3-16
\nonumber \\ &&   \text{sw}^2)) m_z^2) m_t^2+s t^2 (8 t \text{sw}^2+(3-8 \text{sw}^2)
   m_z^2+s (8 \text{sw}^2-6))) \kappa _g \kappa _z m_z^2-s (t-m_t^2)^2
\nonumber \\ &&   ((2 (3-4 \text{sw}^2)^2 m_z^2 (m_z^2-t) m_t^8+2 (3-4
   \text{sw}^2)^2 m_z^2 (t (s+2 t)-2 (s+t) m_z^2) m_t^6-
\nonumber \\ &&    2 (3-4 \text{sw}^2)^2(s^2+4 t s+t^2) m_z^2 (t-m_z^2) m_t^4+s t (-4 (3-4 \text{sw}^2)^2
\nonumber \\ &&   (s+t) m_z^4+(2 s^2 (3-4 \text{sw}^2)^2+6 t^2 (3-4 \text{sw}^2)^2+3 s (64
\nonumber \\ &&   \text{sw}^4-64 \text{sw}^2+15) t) m_z^2-9 s t^2) m_t^2+s^2 t^2 (2 (32
\nonumber \\ &&   \text{sw}^4-24 \text{sw}^2+9) m_z^4-2 (32 \text{sw}^4-24 \text{sw}^2+9) (s+t) m_z^2+9 s
\nonumber \\ &&   t)) g_{\text{gL}}^2+g_{\text{gR}}^2 (32 \text{sw}^4 m_z^2 (m_z^2-t) m_t^8+32
\nonumber \\ &&   \text{sw}^4 m_z^2 (t (s+2 t)-2 (s+t) m_z^2) m_t^6-32 \text{sw}^4 (s^2+4 t s+t^2)
\nonumber \\ &&   m_z^2 (t-m_z^2) m_t^4+s t (-64 \text{sw}^4 (s+t) m_z^4+(32 s^2 \text{sw}^4+96 t^2
\nonumber \\ &&   \text{sw}^4+3 s (64 \text{sw}^4-32 \text{sw}^2+3) t) m_z^2-9 s t^2) m_t^2+s^2 t^2
\nonumber \\ &&   (2 (32 \text{sw}^4-24 \text{sw}^2+9) m_z^4-2 (32 \text{sw}^4-24 \text{sw}^2+9)
\nonumber \\ &&   (s+t) m_z^2+9 s t))) \kappa _g^2)\},
\end{eqnarray}

\begin{eqnarray}
A&=&\frac{8 \text{ln} \left(-\frac{\beta +1}{\beta -1}\right)}{\beta },  \\
B^+&=&-2 \text{Li2}\left(\frac{-\beta  \left(m_t^2-m_z^2+s\right)+m_t^2+m_z^2-s-2 u}{2 m_t^2-2 u}\right)+2 \text{Li2}\left(\frac{\frac{m_t^2+m_z^2-s-2 u}{m_t^2-m_z^2+s}+\beta }{\beta
   -1}\right)+ \nonumber \\ &&
   \text{ln} ^2\left(-\frac{(\beta -1) \left(m_t^2-m_z^2+s\right)}{2 \left(m_t^2-u\right)}\right)-\frac{1}{2} \text{ln} ^2\left(-\frac{\beta +1}{\beta -1}\right),  \\
B^-&=&-2 \text{Li2}\left(-\frac{-(\beta +1) m_t^2+(\beta -1) m_z^2-s \beta +s+2 u}{2 \left(m_z^2-s-u\right)}\right)+2 \text{Li2}\left(\frac{
   \frac{-m_t^2-m_z^2+s+2 \
   u}{  m_t^2-m_z^2+s}+\beta }{\beta -1}\right)+  \nonumber \\ &&
   \text{ln} ^2\left(\frac{(\beta -1) \left(m_t^2-m_z^2+s\right)}{2 \left(m_z^2-s-u\right)}\right)-\frac{1}{2} \text{ln}
   ^2\left(-\frac{\beta +1}{\beta -1}\right),
\end{eqnarray}
where $\beta=\sqrt{1-\frac{4 s m_t^2}{\left(m_t^2-m_z^2+s\right){}^2}}$.

\begin{figure}[h]
      \begin{center}
     \scalebox{0.5}{\includegraphics*{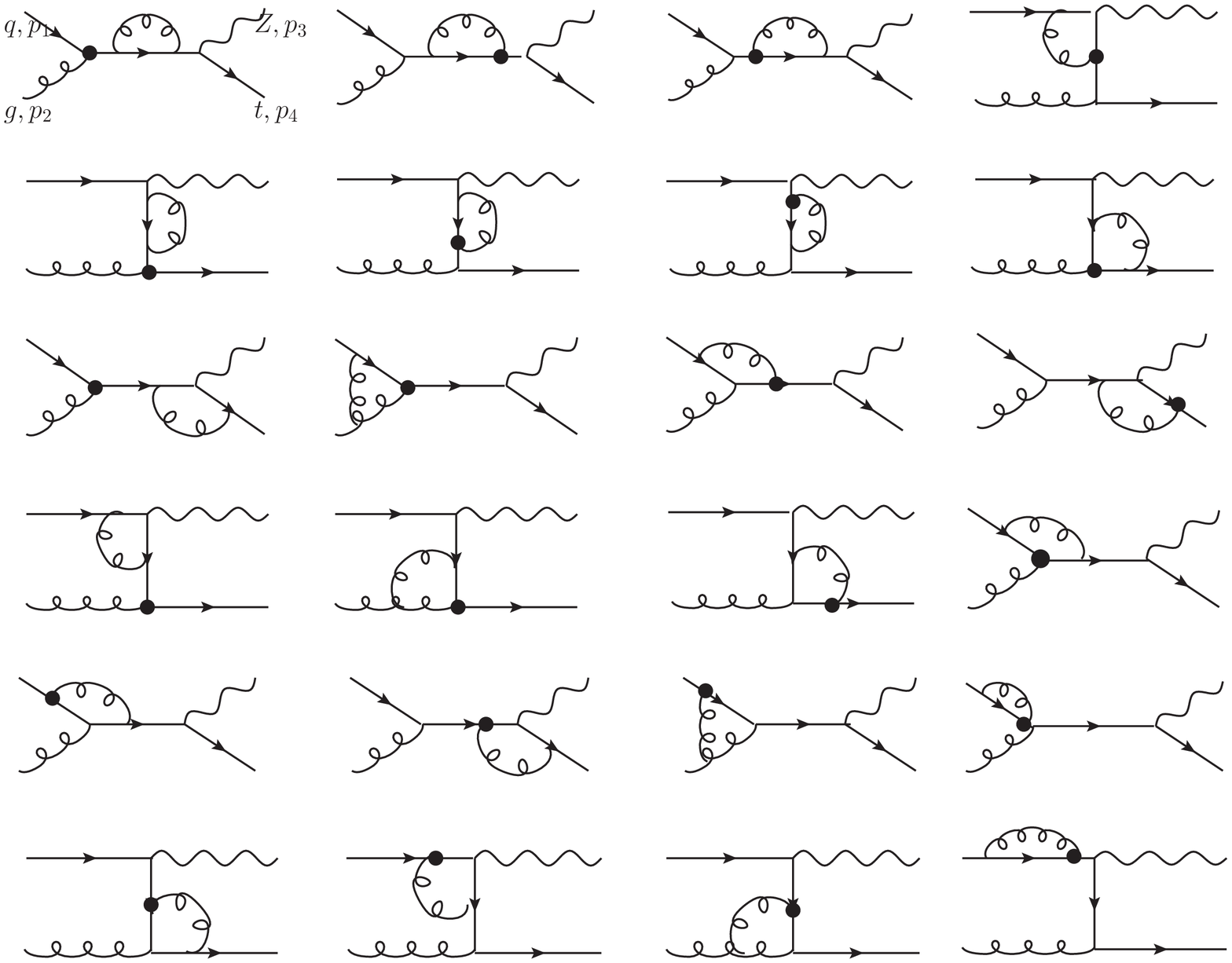}}
      \end{center}
\end{figure}

\begin{figure}[h]
      \begin{center}
     \scalebox{0.6}{\includegraphics*{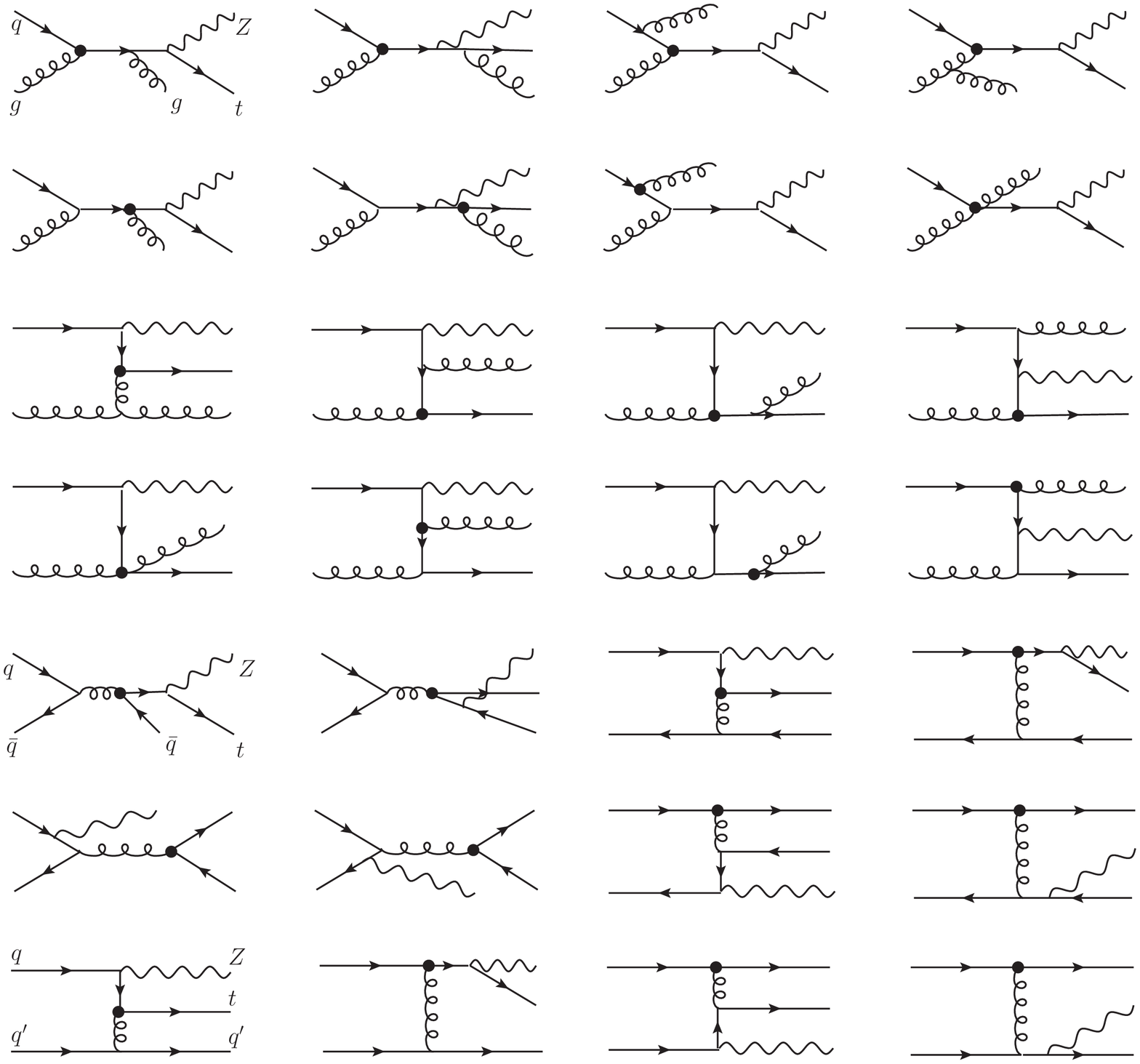}}
      \end{center}
\end{figure}

\end{document}